\documentclass[fleqn,usenatbib]{mnras}

\usepackage{newtxtext,newtxmath}
\usepackage[T1]{fontenc}

\DeclareRobustCommand{\VAN}[3]{#2}
\let\VANthebibliography\thebibliography
\def\thebibliography{\DeclareRobustCommand{\VAN}[3]{##3}\VANthebibliography}

\usepackage{graphicx}	% Including figure files
\usepackage{amsmath}	% Advanced maths commands

\usepackage{amssymb}	% Extra maths symbols
\usepackage[ruled]{algorithm2e}
\usepackage{stfloats}
\usepackage{float}

\usepackage{hyperref}
\usepackage{multirow}
\usepackage{color}

\title[Classification of {\it Gravity Spy} data with GANs]{On Improving the Performance of Glitch Classification for Gravitational Wave Detection by using Generative Adversarial Networks
}

\author[J. Yan et al.]{
	Jianqi Yan,$^{1,3}$
	Alex P. Leung,$^{2}$\thanks{E-mail: alexpl@hku.hk}
        C. Y. Hui,$^{3}$\thanks{Corresponding Author, Email: huichungyue@gmail.com,cyhui@cnu.ac.kr}
	\\
	$^{1}$Faculty of Innovation Engineering, Macau University of Science and Technology, Avenida Wai Long, Taipa, Macau\\
	$^{2}$Department of Physics, University of Hong Kong, Hong Kong\\
        $^{3}$Department of Astronomy and Space Science, Chungnam National University, Daejeon 34134, Korea
}

\date{Accepted XXX. Received YYY; in original form ZZZ}

\pubyear{2021}

\begin{document}
	\label{firstpage}
	\pagerange{\pageref{firstpage}--\pageref{lastpage}}
	\maketitle
	
	\begin{abstract}
		
Spectrogram classification plays an important role in analyzing gravitational wave data. 
In this paper, we propose a framework to improve the classification performance by using Generative Adversarial Networks (GANs). 
As substantial efforts and expertise are required to annotate spectrograms, the number of training examples is very limited. 
However, it is well known that deep networks can perform well only when the sample size of the training set is sufficiently large. 
Furthermore, the imbalanced sample sizes in different classes can also hamper the performance. In order to tackle these problems, we propose a GAN-based data augmentation framework.  
While standard data augmentation methods for conventional images cannot be applied on spectrograms, we found that a variant of GANs, ProGAN, is capable of generating high-resolution spectrograms which are consistent with the quality of the high-resolution original images and provide a desirable diversity. 
We have validated our framework by classifying glitches in the {\it Gravity Spy} dataset with the GAN-generated spectrograms for training.
We show that the proposed method can provide an alternative to transfer learning for the classification of spectrograms using deep networks, i.e. using a high-resolution GAN for data augmentation instead. 
Furthermore, fluctuations in classification performance with small sample sizes for training and evaluation can be greatly reduced. 
Using the trained network in our framework, we have also examined the spectrograms with label anomalies in {\it Gravity Spy}.
		
	\end{abstract}
	
	\begin{keywords}
         Methods: data analysis -- Techniques: image processing -- Gravitational waves
	\end{keywords}
	
	%%%%%%%%%%%%%%%%%%%%%%%%%%%%%%%%%%%%%%%%%%%%%%%%%%
	
	%%%%%%%%%%%%%%%%% BODY OF PAPER %%%%%%%%%%%%%%%%%%
	
	\section{Introduction}
	
Recent state-of-the-art gravitational wave observatories, including Advanced LIGO \citep{aasi2015advanced}, advanced VIRGO \citep{Acernese_2014} and KAGRA \citep{kagra}, have brought the possibility of exploring our Universe in a regime beyond the reach of electromagnetic wave. 
We are now able to investigate the disturbance of spacetime resulted from the merging processes of stellar black holes and neutron stars. 
This not only allows us to study the nature of these compact objects in a brand-new window, but also enables us to test gravitational theories at the limit of strong gravity.
The basic working principle of a gravitational wave detector is laser interferometry. 
The arrival of gravitational wave upon the interferometer will lead to a path difference in the lasers running in two orthogonal arms. 
For a signal with a strain amplitude of $h\lesssim10^{-21}$ \citep[e.g.][]{PhysRevLett.116.061102}, this causes a path difference smaller than the size of a proton between two arms. 
All these new possibilities in gravitational wave astronomy rely upon the unprecedented sensitivity of the detectors in identifying such small changes. 

Apart from the genuine astrophysical signals, these sensitive detectors also pick up unwanted non-Gaussian noise anomalies which are commonly referred as glitches. 
The high rate of these glitches can contaminate the gravitational wave data by mimicking or obscuring the genuine astrophysical signals. 
In order to attain the full potential of the current observatories, it is crucial to identify these noise signals and remove them from the data. 
Combining efforts from the communities of gravitational physics, human-centered computing, machine learning and citizen science, classification of glitches in LIGO data has been published as the {\it Gravity Spy} data \citep{zevin2017gravity}. 
Because of the transient nature of glitches, spectrogram is a convenient data representation as it provides a visualization of how does the frequency spectrum of the signal varies with time. 
The whole {\it Gravity Spy} dataset is segregated into 20 different glitch classes according to the morphologies of their spectrograms. 
Each class corresponds to a different environmental/engineering origin. 
For example, the {\tt Violin Mode Harmonic} class is related to the resonance of the glass fibers that suspend LIGO's main mirrors. 

Since identifying and pinpointing the nature of glitches are important for preventing false detections and improving the instrumental performance, machine learning algorithms have been employed to automate the classification and improve the accuracy (see Section 2.1 for a brief review).
In this work, we present a new architecture to improve the glitch classification performance. There are two main parts in our architecture, include data augmentation and classification.

Apart from the small sample size of training data, the imbalanced classes in the {\it Gravity Spy} dataset (i.e. the numbers of glitches in different classes vary significantly) can also hamper the classification by deep network. 
For example, while there are 7476 spectrograms of {\tt Blip} class in the {\it Gravity Spy} dataset \footnote{{\it Gravity Spy} dataset (version 1.0) is used in all of our experiments: \url{https://zenodo.org/record/1476156\#.YD8BFOgzaUn}}, we have only 108 spectrograms in the {\tt Paired\_doves} class. 
For classification with machine learning techniques, imbalanced data often leads to degradation in generalization performance \citep{sun2009classification,chawla2004special, mahani2019classification}. 
Furthermore, with small sample sizes, it is well known that deep networks do not generalize well and potentially overfit. 

While data augmentation provides a way to tackle the aforementioned problems \citep{tanaka2019data}, traditional approaches on image data (e.g. center crop, random crop, flip, rotation, etc.) cannot ensure the consistency and integrity of the information in the spectrogram. 
For images of celestial objects, such traditional data augmentation methods can exploit certain symmetries (e.g. rotation, translation, reflection) in the data because classification should be invariant to these operations. 
For example, the classification of a galaxy should not be affected by its orientation, location and size in an image \cite[e.g.][]{10.1093/mnras/stv632}. 
However, as each dimension of a spectrogram represents a different quantity (i.e. frequency and time in $y$ and $x$ axes respectively), spectrograms generally lacks the aforementioned symmetries. 
For example, a translation in the frequency dimension of the {\tt Violin Mode Harmonic} data can lead to misclassification as its spectrograms appear as short and dot-like at well-defined frequencies (500 Hz and its integral multiples). 
Therefore, the data augmentation techniques on conventional imaging data cannot be applied to spectrograms. 

In this work, we propose a data augmentation method based on Generative Adversarial Networks (GANs) to generate high-resolution spectrograms for training. 
This method has some successful applications of improving generalizability in the other fields, such as medical imaging \citep[e.g.][]{cyclegan}.
The community of gravitational wave astronomy has also started employing GANs in their works \citep[e.g.][]{McGinn_2021,lopez2022simulating}.
\cite{McGinn_2021} have presented the first application of GANs in generating gravitational wave data. 
By interpolating and sampling in the abstract latent space, the authors have used GANs to explore the space among five classes of simple parametrized waveforms (e.g. sine-Gaussian) to construct new hybrid unmodelled waveforms. 
They further demonstrated that training the detection algorithm on broader range of signals generated by GANs can enhance the performance. 
On the other hand, a GAN-based method has recently been utilized to simulate a glitch class of the largest sample size (i.e. {\tt Blip}) in time domain \citep{lopez2022simulating}, which demonstrates that GANs can generally learn the underlying distribution of this particular class. 

Both of the aforementioned works have applied GANs in generating waveforms in time domain. 
In comparison with the analysis in the time domain, spectrograms have an advantage to better capture the intermittent changes of the signals in both amplitude and frequency. 
By combining the merits in both time and frequency domains, it enables a visualization for the relationships among the temporal, frequency and amplitude variations in the signal. 
This provides us with the motivation to explore the capability of generating spectrograms for all known classes of glitches with GANs.

For conventional GANs, they can only generate low-resolution images as they can only handle resolutions at around 128 $\times$ 128 and 256 $\times$ 256 \citep{miyato2018spectral, liu2017coverless}.
A number of different methods of GANs have been considered and experimented. 
We have found that a variant known as ProGAN, which is recently proposed to generate high-resolution images, works well for our problem with gravitational wave data. 
Using it to learn in the {\it Gravity Spy} dataset, high-resolution spectrograms for each glitch class can be generated with outstanding quality. 
This provides a way to tackle the imbalanced data problem and improve the overall classification accuracy altogether.
Furthermore, our method can avoid overfitting with an augmented training set large and diverse enough for generalization.

This paper is organized as follows: we first give a review on the related works of glitch classification methods and introduce the background for GANs in Section 2. 
In Section 3, datasets used in this work and the procedures for data preprocessing are described. 
The details of our proposed framework are given in Section 4. 
Evaluation criteria and experimental results are presented in Section 5. 
Finally, a summary and discussions are given in Section 6.

	\section{Background}
	
	\subsection{Previous Work}
	
\cite{bahaadini2017deep} have introduced the ``merged-view'' model for glitch classification. 
Each glitch in the {\it Gravity Spy} dataset has four spectrograms of different durations (0.5s, 1.0s, 2.0s, and 4.0s). 
For each event, \cite{bahaadini2017deep} combined four $m\times k$ spectrograms of different duration side by side into a single $2m\times 2k$ spectrogram. 
This approach achieves 96.89$\%$ classification accuracy. 
In later works, ensemble classifier is used to classify the {\it Gravity Spy} dataset \citep{bahaadini2018machine}. 
They achieve an accuracy of 98.21$\%$ on a standard test set . 
In their work, the linear support vector machine (SVM) \citep{mertsalov2009document}, the kernel SVM and the convolutional neural network (CNN) \citep{lawrence1997face} are employed on the data with all four durations. 
Their experimental result first indicates that training with 1.0s duration data achieve higher accuracy than with the other durations.
Then, with the ``merged-view'' model, an accuracy of 97.67$\%$ was achieved in their work.
And by using the ultimate ensemble classifier with hard fusion, an accuracy of 98.21$\%$ was reported.

The method of deep transfer learning \citep{yosinski2014transferable} has been applied in classifying various astronomical images such as galaxies and stellar clusters \citep[e.g.][]{dominguez2019transfer, wei2020deep, khan2019deep}. 
For classifying the spectrograms of glitchs in the {\it Gravity Spy} dataset, \cite{george2017deep,george2018classification} have attempted to employ the technique of transfer learning so as to enhance the performance.
Instead of concatenating spectrograms as \cite{bahaadini2017deep}, the authors adopt a new approach of encoding spectrograms with different durations into an RGB image as an input to specially designed CNNs.
Using the ResNet50 network architecture \citep{he2016deep} and the Inception-V3 network architecture \citep{szegedy2016rethinking}, they have achieved an accuracy of 98.84$\%$ on the test set.
Another novel structure for transfer learning based on Tsallis entropy on the {\it Gravity Spy} dataset is proposed by \cite{ramezani2021transfer}. 
However, this approach only achieves an accuracy of 96.90$\%$.

Although transfer learning apparently improves the performance on a small test set by providing better tuning on the weights, it is uncertain whether it can be well generalized to larger and more diverse data.
The key to a successful knowledge transfer depends on two factors: (a) there is a large in common between the source and target domains, and (b) the target learner is not negatively affected by the transferred knowledge  \citep{zhuang2020comprehensive}.
However, as aforementioned, there is a fundamental difference between the source domain (i.e. ImageNet) and the target domain (i.e. spectrograms) in our case.
Therefore, the uncertainty whether negative transfer will adversely affect the model generalizability when applied to a larger data set can be a concern. 
This provides motivation to explore alternative method for employing deep learning to classify gravitational wave spectrograms.

	\subsection{Generative Adversarial Networks (GANs)}
	
The concept of GANs is proposed by \cite{goodfellow2014generative}. 
As an alternative framework, GANs estimate generative models via an adversarial process which includes a generative model $G$ and a discriminative model $D$. 
The generative model $G$ is used for capturing the data distribution, while the discriminative model $D$ is used to determine whether the sample comes from the data distribution or the model distribution.
	
GANs use backpropagation and dropout algorithm in training $G$ and $D$. 
The training of $D$ and $G$ can be represented by playing a two-player minimax game with value function $V(D, G)$ as given by Equation (\ref{Eq.1}):
	
	\begin{equation}
		\label{Eq.1}
		\resizebox{.93\hsize}{!}{$
		\mathop{min}\limits_{G} \mathop{max}\limits_{D} V(D,G) = E_{x \sim p_{data}(x)} \left[\log D(x)\right] + E_{z \sim p_{z}(z)} \left[\log (1-D(G(z)))\right] $}
	\end{equation}
	
\noindent where $p_{data}$ is the generator's distribution over data $x$, and $p_{z}(z)$ is a prior on input noise variables. 
A mapping to the data space as $G(z;\theta_g)$, where $G$ is a differentiable function represented by a multilayer perceptron with parameters $\theta_g$ is represented. 
$D(x)$ represents the probability that $x$ came from the data.
	
GANs have been demonstrated to be capable in achieving good results on the MNIST \citep{lecun1998gradient}, Toronto Face Database (TFD) \citep{susskind2010toronto}, and CIFAR-10 \citep{krizhevsky2009learning} datasets. 
Shortly thereafter, the conditional generative adversarial nets (CGANs) is presented \cite{mirza2014conditional}. 
They point out that an unconditioned generative model cannot control modes of the generated data. 
Thus, by conditioning the model on additional conditions, it is possible to direct the data generation process.

Based on retaining the original $G$ and $D$, some additional conditions $y$ (e.g. any class labels or other mode data) are combined with the prior input noise $p_z(z)$ in a joint hidden representation in the generator. 
Similarly, the training process of $G$ and $D$ can be represented by Equation (\ref{Eq.2}):
	
	\begin{equation}
		\label{Eq.2}
		\resizebox{.93\hsize}{!}{$ 
			\mathop{min}\limits_{G} \mathop{max}\limits_{D} V(D,G) = E_{x \sim p_{data}(x)} \left[\log D(x|y)\right] + E_{z \sim p_{z}(z)} \left[\log (1-D(G(z|y)))\right] $}
	\end{equation}

Many variants of GANs have been developed in recent years. 
Such as DCGAN \citep{radford2015unsupervised} and WGAN \citep{arjovsky2017wasserstein}. 
DCGAN combines CNN and GANs, and uses full convolution network and batch normalization \citep{ioffe2015batch}. 
WGAN introduces the Wasserstein distance (also known as the Earth-Mover distance) instead of two similarity measures: the Kullback-Leibler divergence (also known as Relative Entropy) and the Jensen-Shannon divergence. 
The Wasserstein distance is used to measure the distance between two distributions.
	
Some variants of GANs are used to generate high-resolution images. 
A general-purpose framework is proposed to predict pixel by pixel \citep{isola2017image}. 
It generates new images with 512 $\times$ 512 resolutions according to input images. 
To tackle the difficult problem of generating high-resolution and realistic images by GANs, a new approach that can produce high-resolution images from semantic label maps is proposed by \cite{wang2018high} which generates 2048 $\times$ 1024 visually appealing results. 
The progressive growing of GANs (ProGAN), which can generate high resolution 1024 $\times$ 1024 face images, is designed by \cite{karras2017progressive}. 
Instead of training all layers of the generator and the discriminator as usual, ProGAN gradually grows new layers to generate high-resolution images.
In this study, we explore the applicability of ProGAN in improving the performance of spectrogram classification.
	
	\section{Data Pre-Processing}
	
Spectrograms in the {\it Gravity Spy} dataset are obtained by Q-transform of transient events in the LIGO time series. 
Each of them is labelled with a unique class \citep{chatterji2004multiresolution}. 
Properties of different classes of glitches in the {\it Gravity Spy} dataset are explained in details by \cite{bahaadini2018machine}.
There are 22 glitch classes in {\it Gravity Spy}. 
The names of these classes are mostly self-explanatory \citep{bahaadini2018machine}. 
Among them, the class {\tt None\_of\_the\_above} collects all spectrograms which cannot be unambiguously assigned to the other 21 classes.  
Also, one should recognize that spectrograms of a number of different classes have rather similar morphologies (e.g. {\tt Light\_Modulation} and {\tt Low\_frequency\_burst}), which can possibly lead to misclassification.
	
\begin{figure}
		\centering
		\includegraphics[scale=0.4]{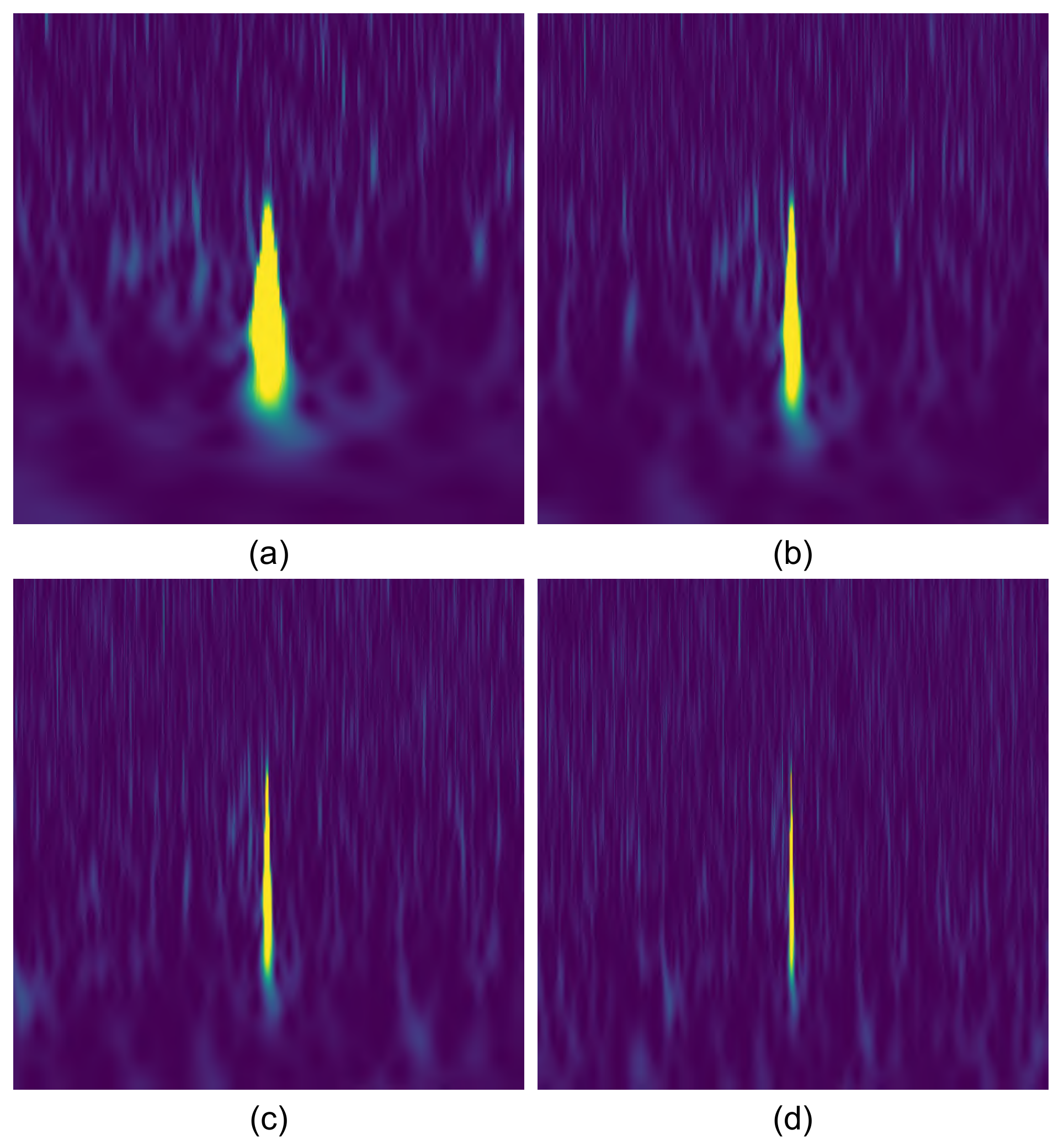}
		\caption{Example spectrograms for the same event in the {\tt Blip} class with different durations: (a) 0.5s, (b) 1.0s, (c) 2.0s, and  (d) 4.0s.}
		\label{fig.3.1}
	\end{figure}

\begin{figure}
		\centering
		\includegraphics[scale=0.3]{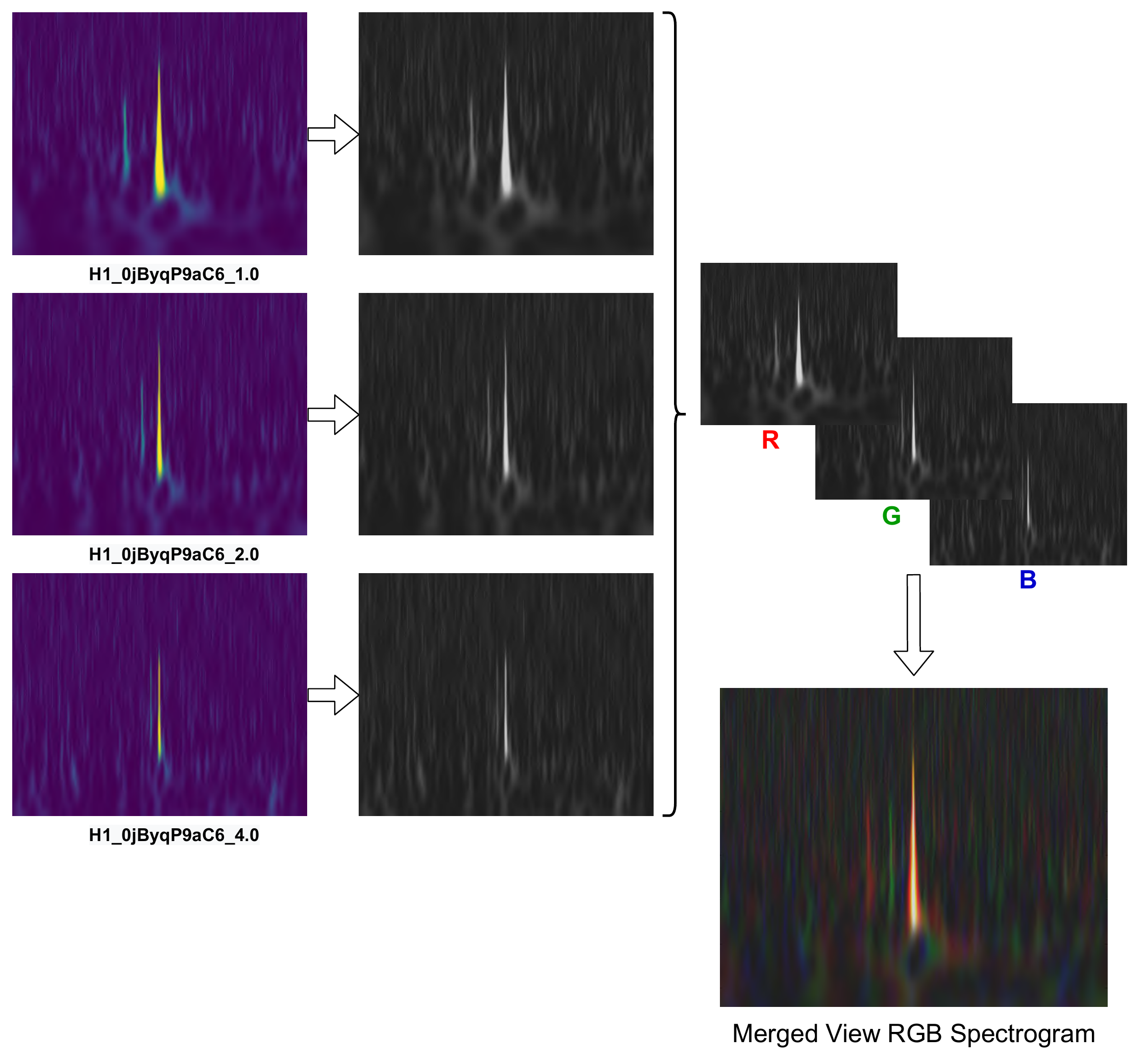}
		\caption{Illustration on how spectrograms of the same event with three different durations are merged into a single RGB spectrogram.}
		\label{fig.3.4}
	\end{figure}

For each glitch, spectrograms with four different durations (0.5s, 1.0s, 2.0s, and 4.0s) are included in the {\it Gravity Spy} dataset. 
Each of them covers a frequency range from 10 Hz to 2048 Hz. 
As an example, we show the spectrograms with different durations of an event of {\tt Blip} class in Figure~\ref{fig.3.1}. 
Following \cite{george2017deep}, we adopted a method of combining three grey-scale spectrograms with three different durations (1.0s, 2.0s and 4.0s) into a single RGB image. 
The method is illustrated by Figure~\ref{fig.3.4}. 

In this work, we have applied the aforementioned data pre-processing on the {\it Gravity Spy} dataset (Version 1.0) as in \citet{george2017deep}.

	\begin{figure}
		\centering
		\includegraphics[scale=0.5]{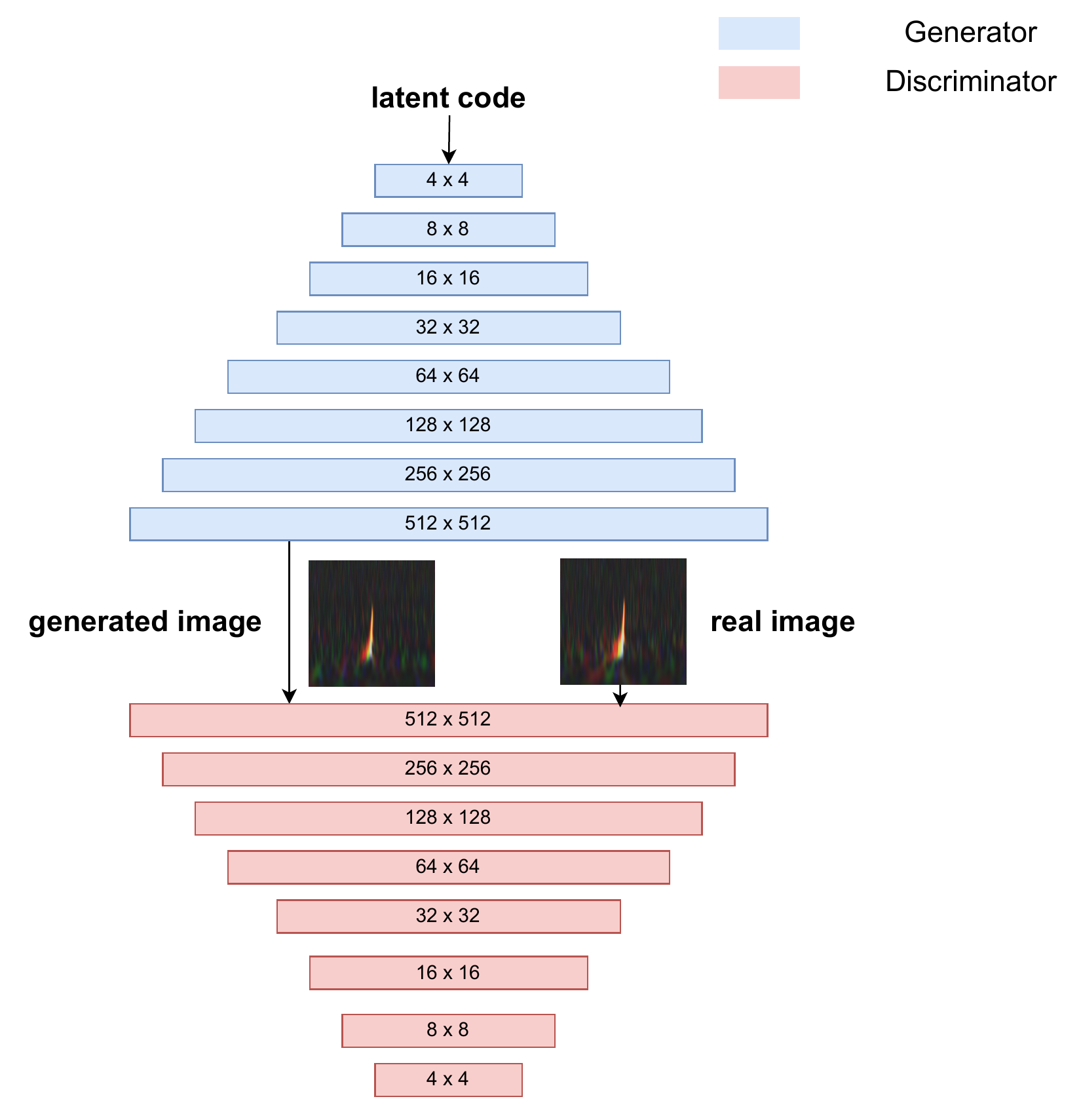}
		\caption{Illustration of the generator and the discriminator of ProGAN. ProGAN generates images with a very low resolution (4 $\times$ 4) in the first layer before the second layer is added to the network architecture for a higher resolution (4 $\times$ 4 to 8 $\times$ 8). This is done recursively. The generator and discriminator always grow in synchrony.}
		\label{fig.4.1}
	\end{figure}
	
	\begin{figure}
		\centering
		\includegraphics[scale=0.4]{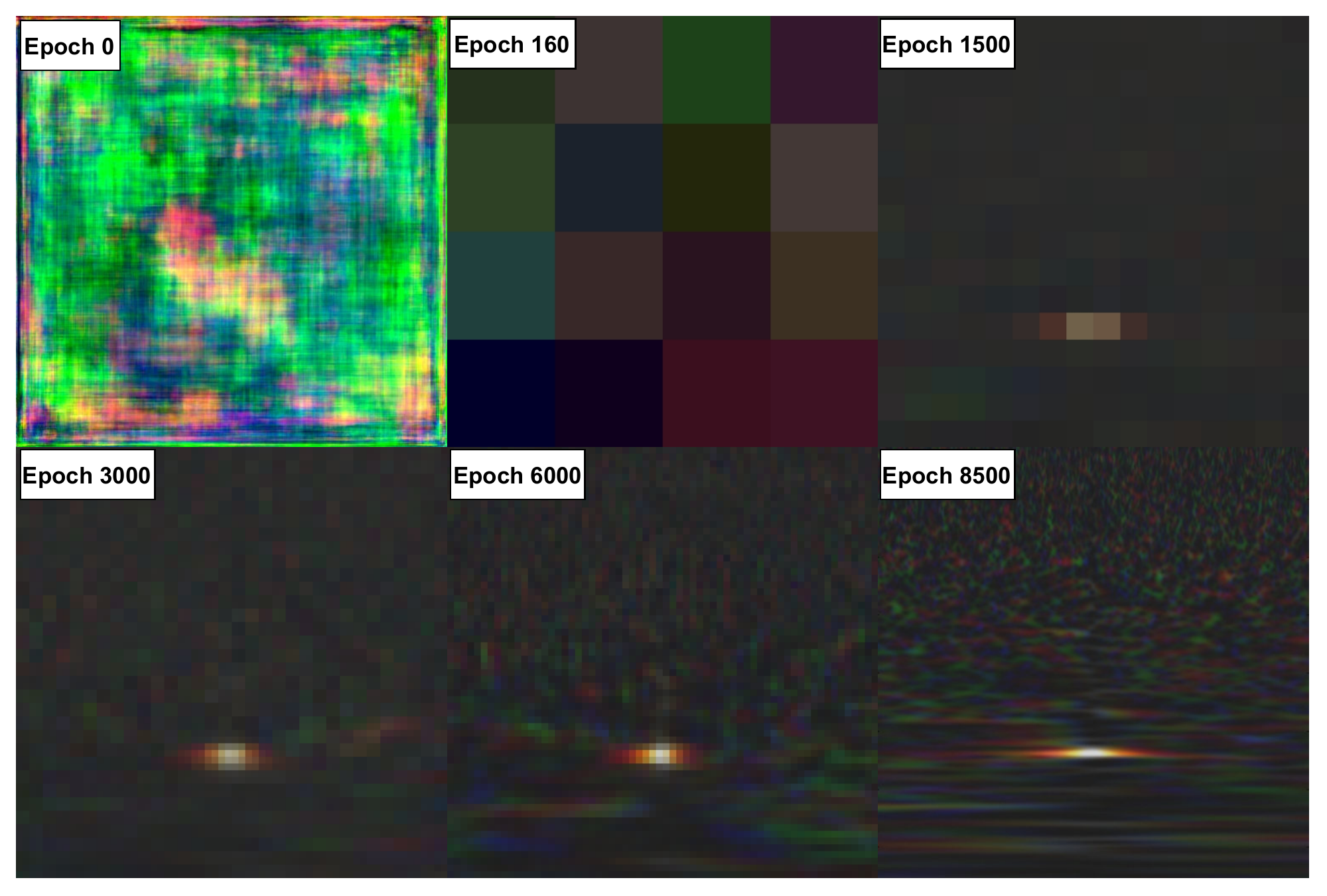}
		\caption{Training process of ProGAN illustrated with an example in the {\tt Air\_Compressor} class. Different epochs of training are shown in a sequential order from left to right and top to bottom. One can notice the progressive increase in resolution.}
		\label{fig.4.2}
	\end{figure}
	
	\begin{figure*}
		\centering
		\includegraphics[scale=0.8]{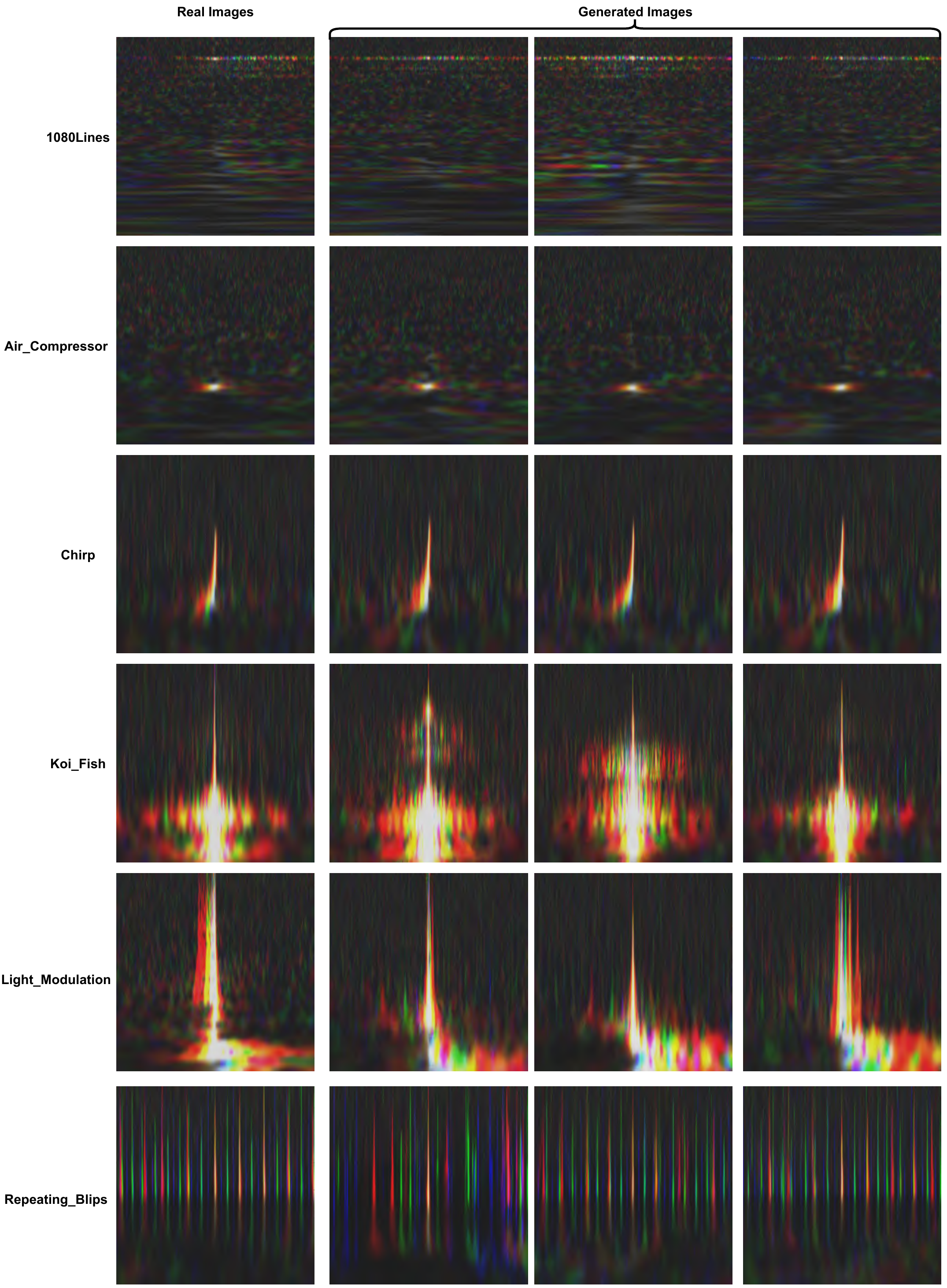}
		\caption{Comparison between original {\it Gravity Spy} spectrograms (Column 1) and the GAN-generated data (Column 2, Column 3, Column 4).}
		\label{fig.4.3}
	\end{figure*}
	
	\begin{table*}
		\centering
		\caption{Different mini-batch configurations for each resolution to minimize the impact of memory constraints in ProGAN.}
		\begin{tabular}{ccccccccc}
			\hline
			& 4$\times$4 & 8$\times$8 & 16$\times$16 & 32$\times$32 & 64$\times$64 & 128$\times$128  &  256$\times$256  &  512$\times$512\\
			\hline
			\textbf{mini-batch} & 128  & 128  & 128  & 128 & 64 & 32 & 16 & 8 \\
			\hline
		\end{tabular}
		\label{table.4.1}
	\end{table*}

	\begin{table*}
		\centering
		\caption{The number of each glitch class in the training set, the validation set, and the test set.}
		\begin{tabular}{lccccc}
			\hline
			Class Name   & Total (V1.0) & multi-duration (RGB) & training set & validation set & test set \\
			\hline
			{\tt 1080lines} &  1312 & 328 & 230 & 49 & 49 \\
			{\tt 1400ripples} & 928 & 232 & 162 & 35 & 35 \\
			{\tt Air\_compressor} & 232 & 58 & 41 & 8 & 9 \\
			{\tt Blip} & 7476 & 1869 & 1308 & 281 & 280   \\
			{\tt Chirp} & 264 & 66 & 46 & 10 & 10         \\
			{\tt Extremely\_loud} & 1816 & 454 &318& 68&68\\
			{\tt Helix} & 1116 & 279 & 195 & 42 & 42      \\
			{\tt Koi\_fish} & 3320 & 830 & 581 & 125 & 124\\
			{\tt Light\_modulation} &2292 & 573 &401&86&86\\
			{\tt Low\_frequency\_burst} & 2628 & 657 & 460 &99&98\\
			{\tt Low\_frequency\_lines} & 1812 & 453 & 317 &68&68\\
			{\tt No\_glitch} & 724 & 181 & 127 & 27 & 27  \\
			{\tt Paired\_doves} & 108 & 27 & 19 & 4 & 4\\
			{\tt Power\_line} & 1812 & 453 & 317 & 68 & 68\\
			{\tt Repeating\_blips} & 1140 & 285 & 200 & 43 &42  \\
			{\tt Scattered\_light} & 1836 & 459 & 321 & 69 & 69 \\
			{\tt Scratchy} & 1416 & 354 & 248 & 53 & 53 \\
			{\tt Tomte} & 464 & 116 & 81 & 17 & 18 \\
			{\tt Violin\_mode} & 1888 & 472 & 330 & 71 & 71 \\
			{\tt Wandering\_line} & 176 & 44 & 31 & 6 & 7 \\
			{\tt Whistle} & 1220 & 305 & 213 & 46 & 46\\
			{\tt None\_of\_the\_above} & 352 & 88 & - & - & - \\
			\hline
		\end{tabular}
		\label{table.4}
	\end{table*}
	
	\section{Our Proposed Framework}
	
	\subsection{Data augmentation}

In this work, we use GANs for data augmentation of spectrograms for gravitational wave data. 
Conventional GANs have difficulties in generating high-resolution spectrograms as they can only handle resolutions at around 128 $\times$ 128 and 256 $\times$ 256 \citep{miyato2018spectral,liu2017coverless}. 
We have investigated different variants of GANs for generating high-resolution spectrograms and experimentally found that with ProGAN, encouraging results can be achieved. 

In the architecture of ProGAN, both generator and discriminator grow progressively: starting from a low resolution (4$\times$4), new layers in a higher resolution (8$\times$8) are then added. 
Increasingly finer details are modelled as the training progresses with more layers. 
This procedure is repeated until the resolution reaches 512$\times$512 or 1024$\times$1024. 
In other words, instead of training all the layers of the generator and the discriminator all at the same time like any other previous GANs approaches, ProGAN gradually increases the number of layers one at a time in order to incorporate high-resolution images. 
	
Figure \ref{fig.4.1} illustrates the structure of ProGAN. 
All generated images from ProGAN and real images are used to train the discriminator. 
ProGAN \footnote{\url{https://github.com/tkarras/progressive_growing_of_gans}} is built on TensorFlow in Python. 
70\% of the training set is used to train ProGAN.
Spectrograms in the {\it Gravity Spy} data are rescaled from the original 566$\times$466 resolution to the 512$\times$512 resolution to adapt to the input resolution of ProGAN.
We set different mini-batches for each resolution to minimize the impact of memory constraints. 
Mini-batch configurations are shown in Table \ref{table.4.1}. 
When the resolution is larger than 32$\times$32, the mini-batch decreases with the increase of resolution. 
The learning rates of the generator are adjusted dynamically. 
Under the situation that the resolution reaches 256$\times$256, the learning rate of the generator is $1.5 \times 10^{-3}$. 
When the last layer of ProGAN is trained, the learning rate of the generator is $3 \times 10^{-3}$. 
As an example, the training process of ProGAN of the {\tt Air\_Compressor} class is shown in Figure \ref{fig.4.2}. 
We use the same parameters for ProGAN to obtain spectrograms in all 21 classes (excluding the {\tt None\_of\_the\_above} class).

In our experiments, the 1.0s, 2.0s and 4.0s spectrograms of each glitch event in the {\it Gravity Spy} dataset are first converted to grayscale images.
Then, those gray images are merged into an RGB image by encoding each gray image into red, green, and blue channels, respectively.
After all grayscale images are combined to form multi-duration RGB spectrograms, the resultant images are resized to 512$\times$512 to match the input size for ProGAN.
The {\it Gravity Spy} dataset are split into three subsets randomly: 70\% for the training set, 15\% for the validation set, and 15\% for the test set (See Table \ref{table.4}).
In view of their ambiguous nature, spectrograms of the {\tt None\_of\_the\_above} class are excluded entirely in the training sets, validation sets, and test sets.
On the other hand, we will use the optimally trained network in an attempt to classify these spectrograms with unknown nature into the other 21 well-defined classes (See Section 5.3).

A comparison between real images (i.e. the original spectrograms in {\it Gravity Spy} dataset) and generated spectrograms by ProGAN is shown in Figure \ref{fig.4.3}. 
Six classes, including the {\tt 1080Lines} class (Row 1), the {\tt Air\_compressor} class (Row 2), the {\tt Chirp} class (Row 3), the {\tt Koi\_Fish} class (Row 4), the {\tt Light\_Modulation} class (Row 5), and the {\tt Repeating\_Blips} class (Row 6) have been chosen for illustration. 
In general, ProGAN not only recovers the training data distribution very well, but also increases the diversity of spectrograms which is important for resulting in less overfits and better generalization (See Sect. 4.3 for further details).

Using GANs to generate high-resolution images is often a time-consuming process. 
The experiment time of training ProGAN are presented by \cite{karras2017progressive}. 
They trained 1024$\times$1024 images on 8 NVIDIA Tesla V100 GPUs for at least 4 days. 
All of our experiments are trained on an NVIDIA Tesla V100 (32GB). 
Each class of the {\it Gravity Spy} dataset is trained by 3 days with a fixed epoch separately. 
However, the image generation process is fast, taking only 5 minutes to generate 1000 512$\times$512 images.

	\subsection{Classification with deep networks}

Many network architectures have been proposed in recent years, including ResNet50 \citep{he2016deep}, ResNet101 \citep{he2016deep}, Inception-V3 \citep{szegedy2016rethinking} and so on. 
ResNet50 and Inception-V3 have been used for gravitational wave glitch classification in the previous works \citep[][]{george2017deep,george2018classification}.

To demonstrate that using GANs as a data augmentation method can improve the spectrogram classification performance, we set up two baselines with: (a) real images (i.e. original spectrograms in {\it Gravity Spy}) to train the aforementioned networks without any pre-trained model, and (b) real images to train these networks with a fine-tuned deep transfer learning method. 
We adopt the pre-train models which are trained with PyTorch on the {\it ImageNet} dataset\footnote{https://pytorch.org/vision/stable/models.html}.
	 
We adopt neural architecture for each network with the number of neurons for the output layer equal to 21 which correspond to the glitch classes defined in {\it Gravity Spy} dataset (excluding {\tt None\_of\_the\_above}).
Also, we resize the 512$\times$512 spectrograms for fitting the specific size of input layer of a particular network architecture. 
We adopt an input dimension of 299$\times$299 for Inception-V3 and 244$\times$244 for all the other networks respectively.

The early-stopping method and the L2-regularization are used to prevent the training process from overfitting. 
In each training epoch, the loss of training set and validation set will be recorded. 
Once the loss of the validation set is greater than the last record for five consecutive times, the training will stop. 
The dynamic adjustment strategy of the learning rate is used in the training process with an initial learning rate of $10^{-4}$.

For resolving the imbalanced data problem, we use ProGAN to generate the spectrograms for different class with Algorithm 1. 
For each class, if $N_{real}$ real spectrograms is less than a predefined number $N_{total}$, we will generate $N_{total} -  N_{real}$ images for compensation. 
We combine the generated images and the original {\it Gravity Spy} spectrograms for the training set of the classification network.
In our experiments, we have tried different values of $N_{total}$ which is taken to be 500, 2000, 3000, and 5000. 
We would like to examine how does the classification performance vary with this parameter.
	
	\begin{algorithm}
		\caption{Data Augmentation Algorithm}
                \label{algorithm1}
		\KwIn{$N_{real}$ real images in one class}
		\KwOut{$N_{aug}$ generated images using ProGAN in each class}
		\For{each class}{
		\eIf{$N_{real} < N_{total}$}{
			$N_{aug} =  N_{total} -  N_{real}$\\
		}{
			$N_{aug} =  0$\\
		}
		Return $N_{aug}$ generated images\\
		}
		
	\end{algorithm}

\begin{figure*}
		\centering
		\includegraphics[scale=0.8]{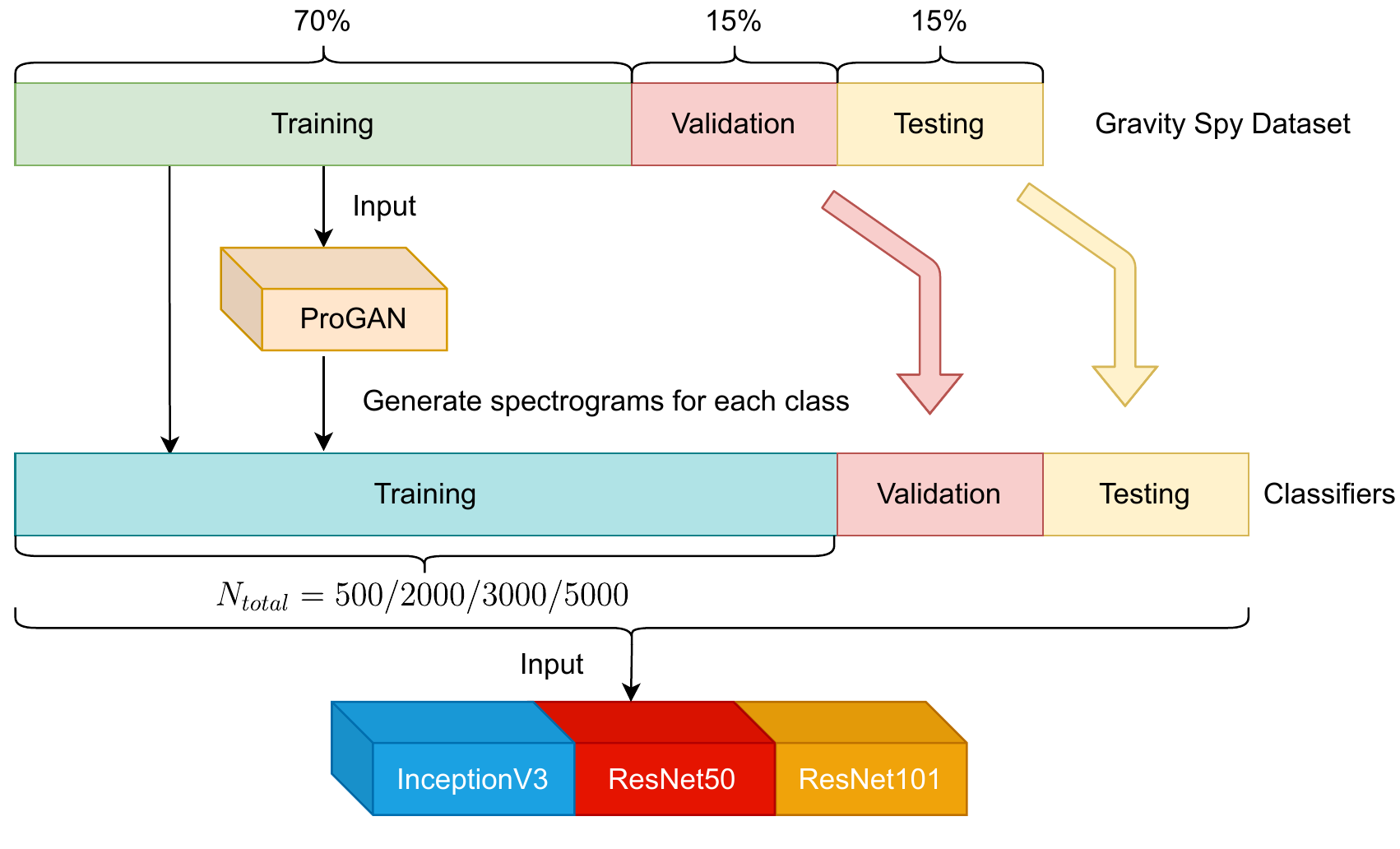}
		\caption{Illustration of our experimental procedure.}
		\label{fig.4.4}
	\end{figure*}

Our experimental procedures are illustrated in Figure \ref{fig.4.4}. 
Prior to splitting the {\it Gravity Spy} dataset, the blank spectrograms in each glitch class are removed. 
The dataset is split into three subsets: 70\% for the training set, 15\% for the validation set, and 15\% for the test set.  
The validation set is used to tune hyperparameters of the network and the test set is used to evaluate the classification performance.
The augmented data by GANs with Algorithm 1 are only added in the training set.
Finally, the confusion matrix and the weighted F1-scores are used to evaluate the glitch classification performance.

In order to evaluate the fluctuations of classification performance, we run each experiment five times. 
During the training process, different layers in network architecture are trained with different random operations (e.g. random weights initialization for the experiments without a pre-trained model, batch normalization, dropout, data shuffling).
On the other hand, we kept the same hyper-parameters initialization (e.g. initial learning rate, batch size, etc.) for all the experiments.
And for those with pre-training incorporated, the weight initialization always starts with the same pre-trained weights in each set of experiments.
	
\subsection{Performance metrics for GANs}	
For evaluating the performance of GANs (i.e. quality of the generated images), there are many effective performance metrics. 
The Inception Score, proposed by \cite{salimans2016improved}, measures the quality of a generated image by computing the KL-divergence between the conditional and marginal probability distributions. 
The Inception Score seeks to capture the image quality and the image diversity of a collection of generated images. 
ProGAN achieves a record Inception Score of 8.80 in unsupervised CIFAR10 \citep{karras2017progressive}. 

However, limitations of the Inception Score is mentioned by \cite{barratt2018note}. 
The authors suggest that the Inception Score is only a good measure for the {\it ImageNet} dataset. 
Furthermore, the Inception Score is sensitive to training weights.
The scores can fluctuate in response to the small changes of the training weights in the same deep learning architecture. 
	
\begin{figure*}
		\centering
		\includegraphics[scale=0.8]{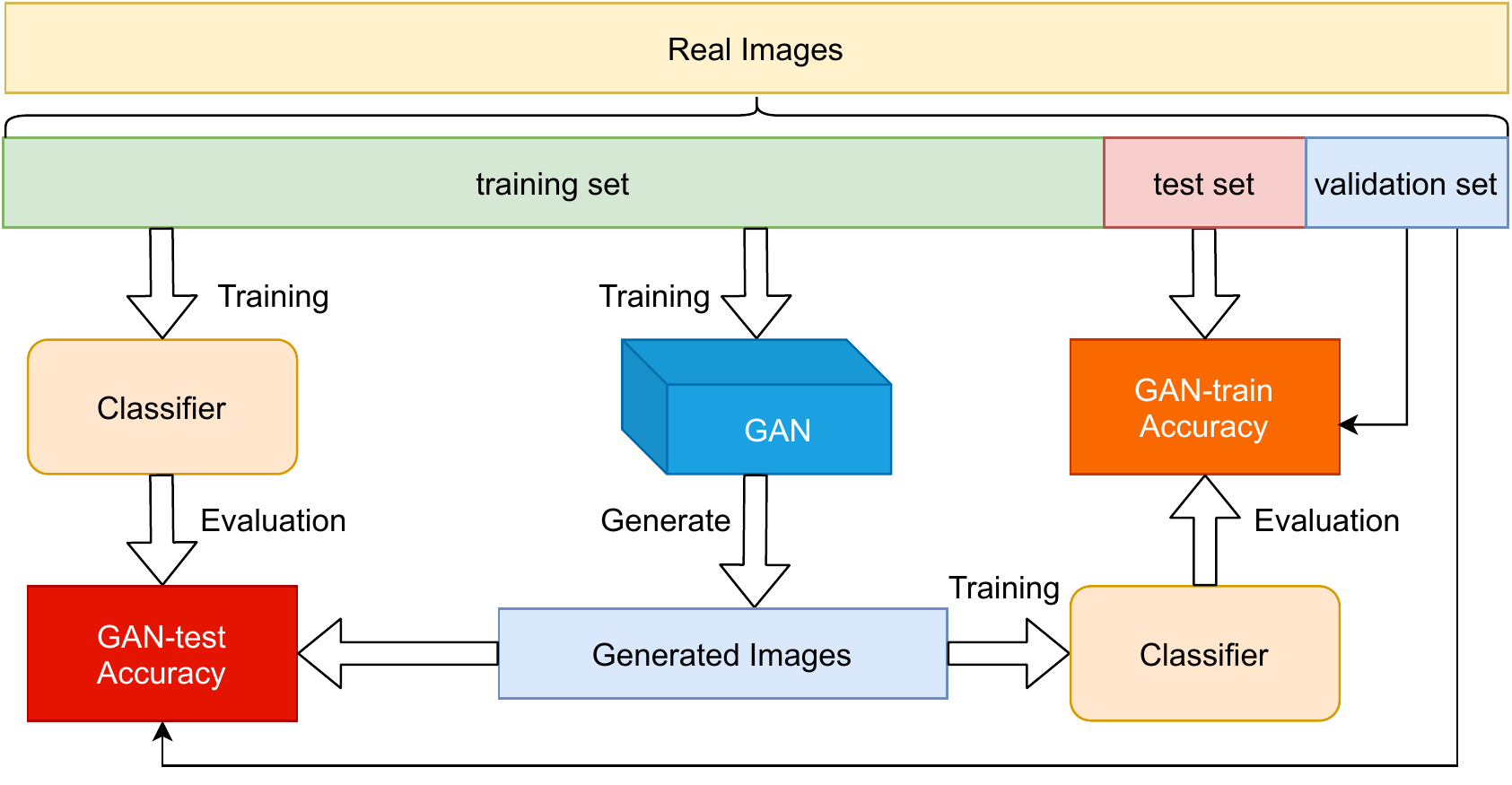}
		\caption{Illustration of the methodology for evaluating the performance of GANs.}
		\label{fig.4.5}
	\end{figure*}

Recently, a novel evaluation method of GANs is proposed by \cite{shmelkov2018good}, which includes two scores: GAN-train and GAN-test. 
While GAN-train learns a classifier on GAN-generated images and measures the performance on real test images, GAN-test measures how realistic GAN-generated images are. 
The procedures are illustrated in Figure \ref{fig.4.5}. 
The diversity and the consistency with the original data of GAN-generated images are evaluated by GAN-train. 
When the GAN-train score on test set is close to that on the validation set, it shows that the diversity of generated images is desirable.
On the other hand, GAN-test learns a classifier on real images and evaluates this classifier on GAN-generated images. 
A high GAN-test means that the quality of generated images is close to the quality of real images. 
Ideally, GAN-test score on the test set should be close to the score of the validation set. 
	
Both GAN-train score and GAN-test score are used to evaluate the performance of ProGAN in our experiments. 
In this work, spectrograms in the {\it Gravity Spy} dataset are randomly split into three subsets: 70\% for the training set, 15\% for the validation set, and 15\% for the test set. 
We generate 10000 spectrograms for each class. 
We use the training set to train a classifier and evaluate the GAN-test score by using the generated images as the test set in this experiment. 
Similarly, we use the generated images to train a new classifier and evaluate the GAN-train score by using the test set. 
In this experiment, the network architectures of classifiers are ResNet50, ResNet101 and Inception-v3.

	\section{Experimental Results}
	
	\subsection{Evaluation of GANs performance in spectrogram generations}

	\begin{table}
		\centering
		\caption{Comparison of GAN-train and GAN-test scores for different network architectures in our experiment.}
		\begin{tabular}{lcccc}
			\hline
			& ResNet50 & ResNet101 & Inception-V3\\
			& (244$\times$244) &  (244$\times$244) &  (299$\times$299)\\
			\hline
			GAN-train &97.88(\%) & 97.64(\%) &  \textbf{98.43}(\%)\\
			GAN-test  &99.31(\%) & \textbf{99.35}(\%) &  95.87(\%) \\
			\hline
		\end{tabular}
		\label{table.5.1}%
	\end{table}
	
	\begin{figure*}
		\centering
		\includegraphics[scale=0.85]{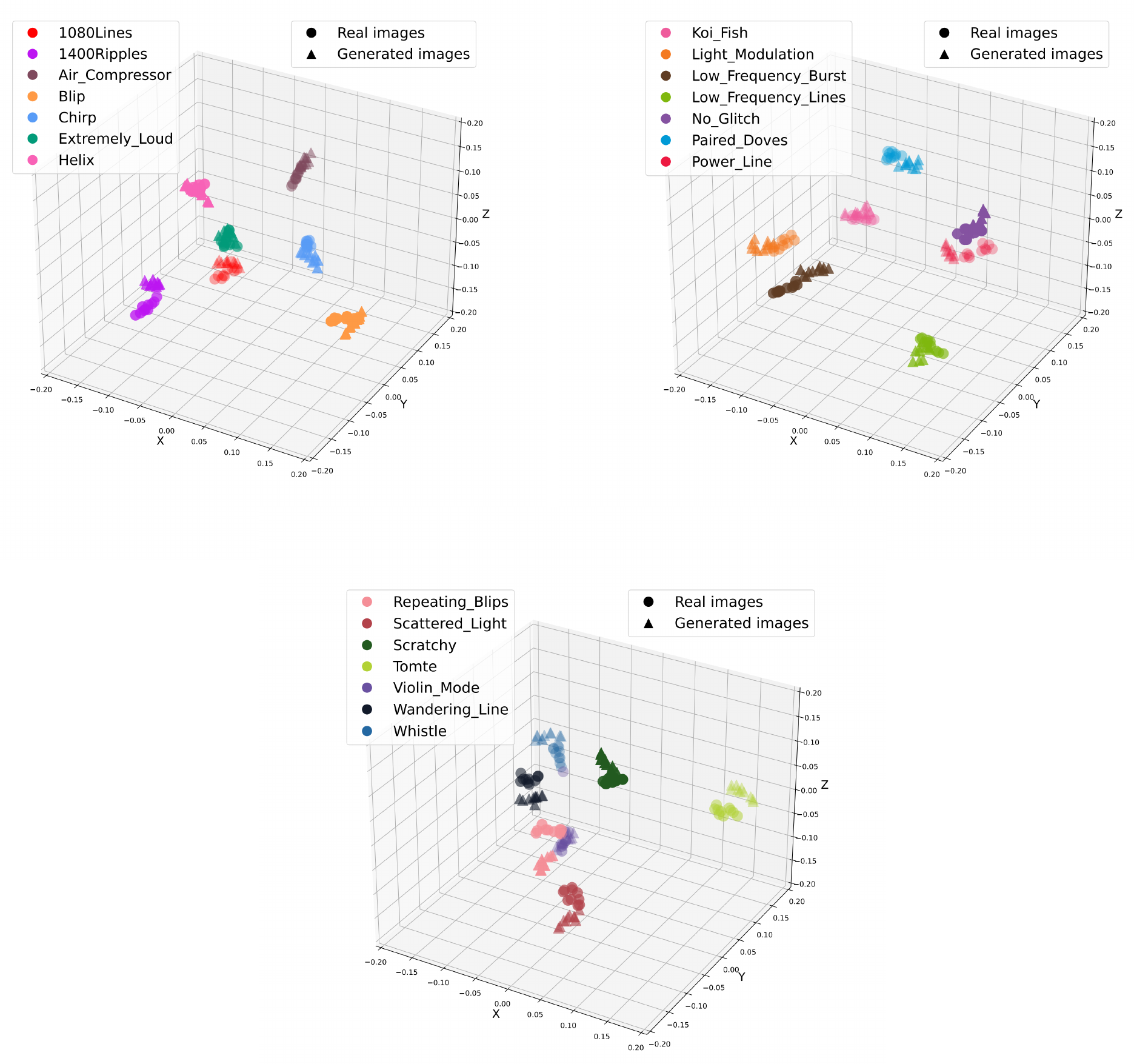}
		\caption{\textbf{Visualization of 21 classes in the {\tt Gravity Spy} dataset after applying clustering and dimensionality reduction by the t-SNE algorithm.}}
		\label{fig.5.1}
	\end{figure*}
	
With 10000 generated images for each class by ProGAN, the performance of different network architectures is compared by using GAN-train and GAN-test scores in Table \ref{table.5.1}. 
In our experiment with multi-duration RGB spectrograms, both GAN-train and GAN-test scores are desirable. 
The highest GAN-train and GAN-test are 98.43\% and 99.35\%, respectively.
This implies that the multi-duration spectrograms generated by GANs are sufficiently diverse and realistic. 
Overall, based on these performance metrics, we conclude that ProGAN is capable to generate images with an acceptable quality and diversity with the multi-duration RGB spectrograms. 

In order to further examine the consistency between GAN-generated images and real images, we use the backbone of the Inception-V3 network architecture for feature extraction and t-SNE \citep{van2008visualizing} for unsupervised learning. 
To validate the effectiveness of our GAN-based framework, we use the backbone of the optimal Inception-V3 
network architecture ($N_{total}$= 5000, Without Pre-training cf. Table \ref{table.5.2}) and remove its final fully connected layer. 
Therefore, for all multi-durations RGB spectrogram inputs, the trained model outputs a high dimensional feature vector and this vector is used as input for t-SNE. 
We used t-SNE to reduce the high-dimensional feature vectors into a 3D feature space for visualization. 
10 real images and 10 GAN-generated images in each class are randomly selected for the experiments. 
 
Figure \ref{fig.5.1} shows the clustering results of the t-SNE algorithm. 
Different colours represent different types of glitches. 
We found that the real images (circles in Figure \ref{fig.5.1}) and the generated images (triangles in Figure \ref{fig.5.1}) from the same class are clustered in a similar way in this reduced 3D feature space. 

We notice that a generated image from the {\tt Violin\_Mode} class is clustered with those from the {\tt Whistle} class (see the bottom panel of Figure \ref{fig.5.1}). 
This may account for a misclassification between these two classes as seen in the confusion matrix shown in Figure \ref{fig.5.3}.  
Except for this, we found the GAN-generated data are generally consistent with the original {\tt Gravity Spy} data. 

For testing the robustness of the clustering results by t-SNE, we have run the experiment with 7 types of glitches for 9 times. 
We have included {\tt 1080Lines}, {\tt 140Ripples}, {\tt Air\_Compressor}, {\tt Blip}, {\tt Chirp}, {\tt Extremely\_Loud} and {\tt Helix} in this robustness test. Different random seeds were set in each run.
The results are shown in Appendix \ref{appendix_A}. 
Both real and generated images are found to be clustered closely and different classes are well differentiated in each experiment. 
In summary, we conclude that the GAN-generated images are correctly clustered with the real data. 

\subsection{Classification of {\it Gravity Spy} data}

Top-1 accuracy is the proportion of correctly classified images with their true classes equal to the predict classes with the highest probabilities (i.e. the conventional definition of accuracy), which is the common evaluation criteria to gauge the classification performance in image classification \citep{simonyan2014very}.
However, top-1 accuracy is most used in balanced data when all the classes are equally important.
On the other hand, there are many works using F1-score to evaluate the classification performance \citep[e.g.][]{li2017improving, sullivan2018deep}.
F1-score is based on the harmonic mean of precision and recall.
The most common F1-scores include the micro-averaged F1-score, the macro-averaged F1-score and the weighted F1-score.
While micro-averaged F1-score is obtained by computing total precision and recall globally, macro-averaged F1-score is the unweighted average of the precision and recall for each class.

The weighted F1-score is a variant of macro-averaged F1-score.
The key difference between the weighted F1-score and the macro-averaged F1-score is how precision and recall are computed for each class.
According to the sample size of the test set and the number of true images of each class, weights are determined for each class and the weighted F1-score is calculated according to the weighted precision and the weighted recall for each class.
In view of the imbalanced classes in the test set, we decided to use the weighted F1-score instead of the top-1 accuracy to evaluate the classification performance in this work.
In addition, we also used the confusion matrix is to visualize the performance in our experiments. 
	
In order to increase the number of spectrograms in each class in the training data and tackle the imbalanced data problem, we mixed the spectrograms generated by ProGAN with the real images from the {\it Gravity Spy} data for training (cf. \autoref{algorithm1}).
In Table \ref{table.5.2}, we compare the weighted F1-score of different $N_{total}$ on different network architectures.
We have run each experiment five times for observing the fluctuation of classification performance (cf. Section 4.2).
The average weighted F1-score of each experiment are shown in Table \ref{table.5.2}. 
The quoted uncertainties correspond to the highest/lowest weighted F1-scores obtained among five runs for each experiment.

First, we set up two baselines ($N_{aug}=0$) as mentioned in Section 4.2 (i.e. one with pre-training and one without pre-training). 
Without any data augmentation methods, we found that the pre-trained model reaches a higher weighted F1-score than the one without pre-training regardless of the network architectures.
This conclusion is consistent with previous findings \citep{george2017deep,george2018classification}.

\begin{figure*}
		\centering
		\includegraphics[scale=0.40]{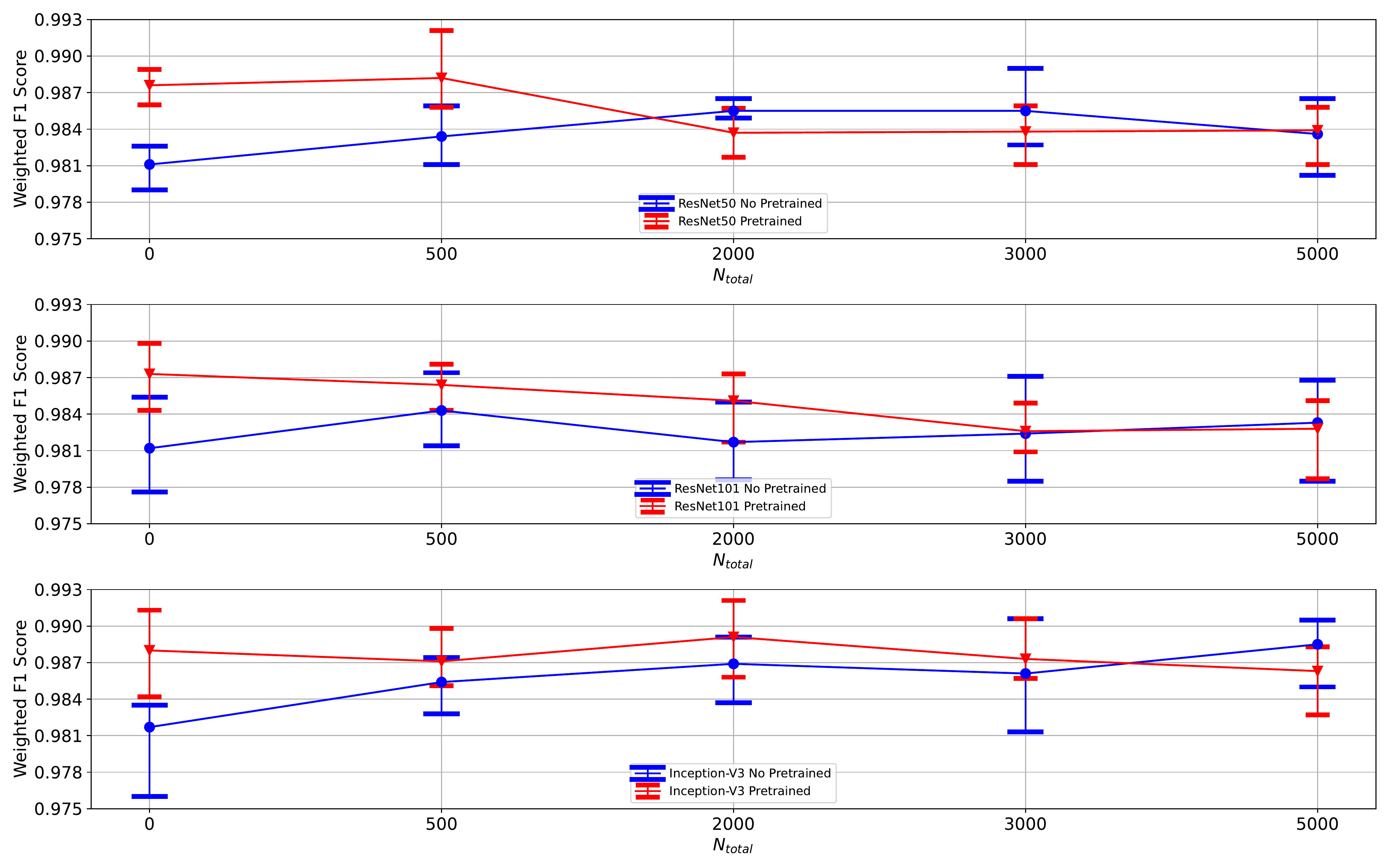}
		\caption{Illustration of how weighted F1 scores vary with $N_{total}$ in different network architectures. The blue lines and symbols show the results from the experiments without pre-training. And the red lines and symbols show the results with pre-trained models. The maximum performance fluctuations among five runs in each experiment are shown with the error bars.}
		\label{fig.4.10}
	\end{figure*}

Once the baselines have been set, we performed a series of experiments with different $N_{total}$ for comparisons. 
The results are summarized in Table~\ref{table.5.2} and Figure~\ref{fig.4.10}. 
The quoted error bars correspond to the maximum variations of weighted F1-scores.
For all the network architectures considered in this work, the GAN-based data augmentation method can generally lead to more accurate classification with respect to the baselines without pre-training. 
In the case of Inception-V3, it achieves the highest average weighted F1-score of 0.9885 for $N_{total}=5000$ without pre-training which is higher than both corresponding baselines.
This shows that our method can lead to a performance improvement comparable or even exceed those attained by transfer-learning.
For the experiments which incorporated both pre-training and data augmentation, the highest average weighted F1-score of 0.9891 is achieved by the Inception-V3 network architecture with $N_{total}=2000$ and pre-training.

	\begin{table*}
		\centering
		\renewcommand\arraystretch{1.3}
		\caption{The average weighted F1 scores of glitch classification on the multi-duration test set from the {\it Gravity Spy} dataset. The maximum fluctuations among five runs in each experiment are shown.} 
		\begin{tabular}{lccc}
			\hline
			\hline
			& ResNet50 (224$\times$224) & ResNet101 (224$\times$224) & Inception-V3  (299$\times$299)  \\
			\hline
			\hline
                      \multicolumn{4}{c}{Baselines (i.e. without data augmentation)}\\
            \hline
			Without pre-training & 0.9811 $\pm ^{\scriptscriptstyle \text{0.0015}}_{\scriptscriptstyle \text{0.0021}}$ & 0.9812 $\pm ^{\scriptscriptstyle \text{0.0042}}_{\scriptscriptstyle \text{0.0036}}$ & 0.9817 $\pm ^{\scriptscriptstyle \text{0.0018}}_{\scriptscriptstyle \text{0.0057}}$ \\
			With pre-training &  0.9876 $\pm ^{\scriptscriptstyle \text{0.0013}}_{\scriptscriptstyle \text{0.0016}}$ & 0.9873 $\pm ^{\scriptscriptstyle \text{0.0025}}_{\scriptscriptstyle \text{0.0030}}$ & 0.9880 $\pm ^{\scriptscriptstyle \text{0.0033}}_{\scriptscriptstyle \text{0.0038}}$ \\
			\hline
                        \multicolumn{4}{c}{With data augmentation only}\\
			\hline
			$N_{total}$ = 500 (Without Pre-training) & 0.9834 $\pm ^{\scriptscriptstyle \text{0.0025}}_{\scriptscriptstyle \text{0.0023}}$ & \textbf{0.9843} $\pm ^{\scriptscriptstyle \text{0.0031}}_{\scriptscriptstyle \text{0.0029}}$ & 0.9854 $\pm ^{\scriptscriptstyle \text{0.0020}}_{\scriptscriptstyle \text{0.0026}}$ \\
			$N_{total}$ = 2000 (Without Pre-training)& \textbf{0.9855} $\pm ^{\scriptscriptstyle \text{0.0010}}_{\scriptscriptstyle \text{0.0006}}$ & 0.9817 $\pm ^{\scriptscriptstyle \text{0.0033}}_{\scriptscriptstyle \text{0.0031}}$ & 0.9869 $\pm ^{\scriptscriptstyle \text{0.0022}}_{\scriptscriptstyle \text{0.0032}}$ \\
			$N_{total}$ = 3000 (Without Pre-training)& 0.9855 $\pm ^{\scriptscriptstyle \text{0.0035}}_{\scriptscriptstyle \text{0.0028}}$ & 0.9824 $\pm ^{\scriptscriptstyle \text{0.0047}}_{\scriptscriptstyle \text{0.0039}}$ & 0.9861 $\pm ^{\scriptscriptstyle \text{0.0045}}_{\scriptscriptstyle \text{0.0048}}$ \\
			$N_{total}$ = 5000 (Without Pre-training)& 0.9836 $\pm ^{\scriptscriptstyle \text{0.0029}}_{\scriptscriptstyle \text{0.0034}}$  & 0.9833 $\pm ^{\scriptscriptstyle \text{0.0035}}_{\scriptscriptstyle \text{0.0048}}$ & \textbf{0.9885} $\pm ^{\scriptscriptstyle \text{0.0020}}_{\scriptscriptstyle \text{0.0035}}$ \\
			
			\hline
			\multicolumn{4}{c}{With both data augmentation and pre-training}\\
			\hline
			$N_{total}$ = 500 & \textbf{0.9882} $\pm ^{\scriptscriptstyle \text{0.0039}}_{\scriptscriptstyle \text{0.0024}}$ & 0.9864 $\pm ^{\scriptscriptstyle \text{0.0017}}_{\scriptscriptstyle \text{0.0021}}$ & 0.9871 $\pm ^{\scriptscriptstyle \text{0.0027}}_{\scriptscriptstyle \text{0.0020}}$ \\
			$N_{total}$ = 2000& 0.9837 $\pm ^{\scriptscriptstyle \text{0.0020}}_{\scriptscriptstyle \text{0.0020}}$ & 0.9851 $\pm ^{\scriptscriptstyle \text{0.0022}}_{\scriptscriptstyle \text{0.0034}}$ & \textbf{0.9891} $\pm ^{\scriptscriptstyle \text{0.0030}}_{\scriptscriptstyle \text{0.0033}}$  \\
			$N_{total}$ = 3000& 0.9838 $\pm ^{\scriptscriptstyle \text{0.0021}}_{\scriptscriptstyle \text{0.0027}}$ & 0.9826 $\pm ^{\scriptscriptstyle \text{0.0023}}_{\scriptscriptstyle \text{0.0017}}$ & 0.9873 $\pm ^{\scriptscriptstyle \text{0.0033}}_{\scriptscriptstyle \text{0.0016}}$ \\
			$N_{total}$ = 5000& 0.9839 $\pm ^{\scriptscriptstyle \text{0.0019}}_{\scriptscriptstyle \text{0.0028}}$ & 0.9828 $\pm ^{\scriptscriptstyle \text{0.0023}}_{\scriptscriptstyle \text{0.0041}}$ & 0.9863 $\pm ^{\scriptscriptstyle \text{0.0020}}_{\scriptscriptstyle \text{0.0036}}$  \\
			\hline
			\hline
		\end{tabular}
		\label{table.5.2}%
	\end{table*}

	\begin{figure*}
		\centering
		\includegraphics[scale=0.429]{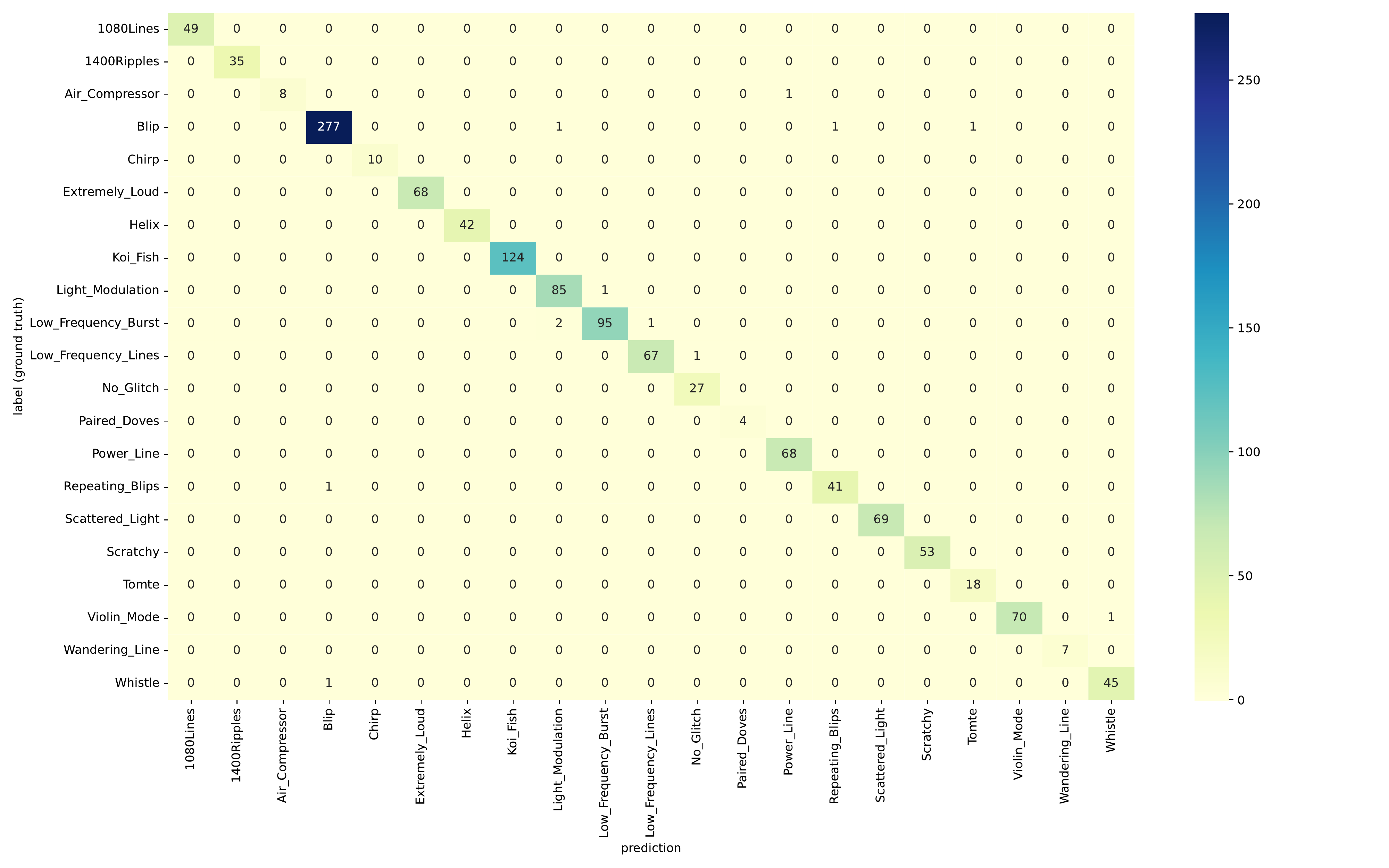}
		\caption{Confusion matrix with experimental results from the Inception-V3 network architecture ($N_{total} = 5000$, without pre-training).}
		\label{fig.5.3}
	\end{figure*}
	
	\begin{figure*}
		\centering
		\includegraphics[scale=0.429]{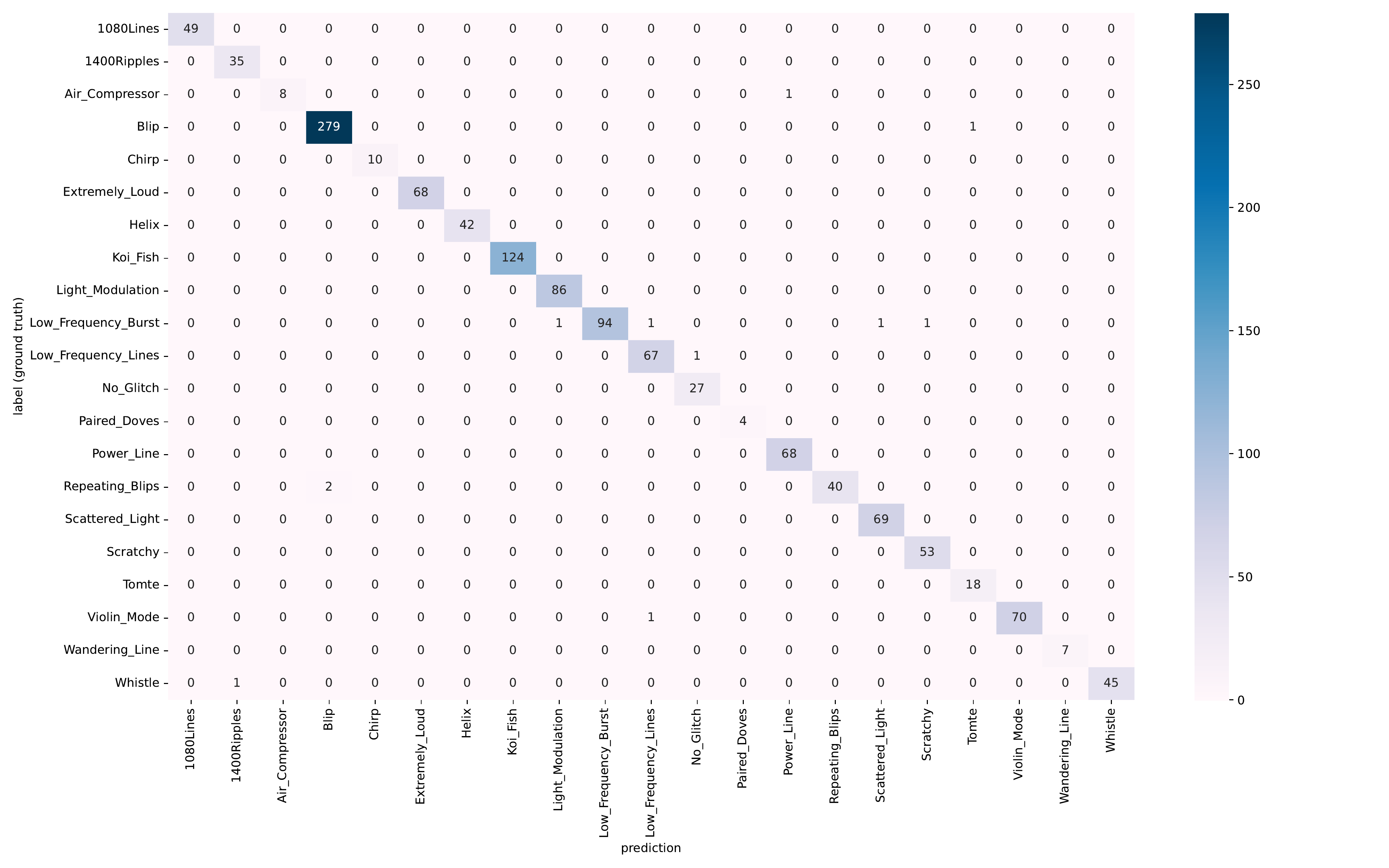}
		\caption{Confusion matrix with experimental results from the Inception-V3 network architecture ($N_{total} = 2000$, with pre-training).}
		\label{fig.5.4}
	\end{figure*}
	
We have also investigated the impact of GAN-based data augmentation on recognizing the classes with small sample size. 
We focus on the classes {\tt Air\_Compressor}, {\tt Paired\_Doves} and {\tt Wandering\_Line}, in the {\it Gravity Spy} dataset.
In this part, we only adopted the Inception-V3 network architecture because of its superior performance as it allows a larger input size.
In Table \ref{table.5.3}, apart from the averaged weighted F1-scores (Column 8) and the standard deviations (Column 9), the weighted F1-scores in each five runs of all the experiments are also explicitly shown in Columns 3-7.

The baseline performances are given in the first two rows of Table \ref{table.5.3}. 
Without data augmentation and pre-training, the classification performance on the small classes apparently suffer from significant fluctuations. 
The largest impact is found on the class with the smallest sample size (i.e. {\tt Paired\_Doves}), which has the weighted F1-score varied from 0.667 to 1 in our experiments. 
On the other hand, for the baseline with pre-training incorporated, the fluctuations are reduced considerably (e.g. perfect precision and recall on {\tt Paired\_Doves} are achieved).

To examine the influence of our framework, we incorporated data augmentation by varying $N_{total}$ without any pre-training. 
The results are summarized in Rows 3-6 of Table \ref{table.5.3}. 
We found that the fluctuations in all three classes can generally be reduced by our method. 
For the case of {\tt Paired\_Doves}, we obtained perfect precision and recall even with $N_{total}=500$. 
For {\tt Wandering\_Line}, we have achieved an average weighted F1-score of 0.987 with $N_{total}=5000$ which exceeds that attained by the baseline with pre-training.

For the case of {\tt Air\_Compressor}, we notice that in all the experiments with pre-training and/or GAN-based data augmentation incorporated, the weighted F1-score kept fluctuating between two values: 0.941 and 1. 
Examining the confusion matrices found that one of the nine samples in this test set is frequently mis-classified in the class of {\tt Power\_Line}. 
This prompts us to inspect its test data visually. 
We found that the spectrogram of this mis-classified data remarkably resembles that of {\tt Power\_Line}. 
We speculate that this might be ascribed to the misclassification in the original {\it Gravity Sky} data.

Apart from the aforementioned problem in the class of {\tt Air\_Compressor}, we found that GAN-based data augmentation can generally improve the weighted F1-scores and reduce their fluctuations in classifying small samples.

	\begin{table*}
		\centering
		\renewcommand\arraystretch{1.3}
		\caption{The weighted F1-scores for classifying the multi-duration test set of three classes with small sample sizes from the {\it Gravity Spy} dataset.}
		\begin{tabular}{lcccccccc}
			\hline
			\hline
			\textbf{}              &     & \textbf{1} & \textbf{2} & \textbf{3} & \textbf{4} & \textbf{5} & \textbf{Average Weighted F1 Score} & \textbf{Standard Deviation} \\
			\hline
			\hline
			\multicolumn{9}{c}{Baselines (i.e. without data augmentation)}\\
			\hline
			\hline
			\multirow{3}{*}{Without Pre-training} & Air\_Compressor&1&0.941&1&1&1&0.988&0.026\\
			& \multicolumn{1}{l}{Paired\_Doves}   & \multicolumn{1}{c}{0.667} & \multicolumn{1}{c}{0.667} & \multicolumn{1}{c}{1} & \multicolumn{1}{c}{0.667} & \multicolumn{1}{c}{0.667} &
			\multicolumn{1}{c}{0.733} &\multicolumn{1}{c}{0.149} \\
			& \multicolumn{1}{l}{Wandering\_Line} & \multicolumn{1}{c}{0.833} & \multicolumn{1}{c}{0.727} & \multicolumn{1}{c}{1} & \multicolumn{1}{c}{0.714} & \multicolumn{1}{c}{0.923} &
			\multicolumn{1}{c}{0.840} &\multicolumn{1}{c}{0.124}  \\
			\hline
			\multirow{3}{*}{Pre-training} & Air\_Compressor&1&1&1&1&0.941&0.988& 0.026\\
			& \multicolumn{1}{l}{Paired\_Doves}   & \multicolumn{1}{c}{1} & \multicolumn{1}{c}{1} & \multicolumn{1}{c}{1} & \multicolumn{1}{c}{1} &  \multicolumn{1}{c}{1} &
			\multicolumn{1}{c}{1} &\multicolumn{1}{c}{0}   \\
			& \multicolumn{1}{l}{Wandering\_Line} & \multicolumn{1}{c}{1} & \multicolumn{1}{c}{1} & \multicolumn{1}{c}{1} & \multicolumn{1}{c}{0.923} & \multicolumn{1}{c}{1} &
			\multicolumn{1}{c}{0.985} &\multicolumn{1}{c}{0.034}  \\
			\hline
			\hline
			\multicolumn{9}{c}{With data augmentation only}\\
			\hline
			\hline
			\multirow{3}{*}{500 (Without Pre-training)} & Air\_Compressor&1&1&1&0.941&0.941&0.978&0.031\\
			& \multicolumn{1}{l}{Paired\_Doves}   & \multicolumn{1}{c}{1} & \multicolumn{1}{c}{1} & \multicolumn{1}{c}{1} & \multicolumn{1}{c}{1} & \multicolumn{1}{c}{1} & \multicolumn{1}{c}{1} & \multicolumn{1}{c}{0} \\
			& \multicolumn{1}{l}{Wandering\_Line} & \multicolumn{1}{c}{0.875} & \multicolumn{1}{c}{1} & \multicolumn{1}{c}{1} & \multicolumn{1}{c}{0.875} & \multicolumn{1}{c}{0.933} & \multicolumn{1}{c}{0.937} &
			\multicolumn{1}{c}{0.063} \\
			\hline
			\multirow{3}{*}{2000 (Without Pre-training)} & Air\_Compressor&1&0.941&1&0.941&1&0.976&0.032\\
			& \multicolumn{1}{l}{Paired\_Doves}   & \multicolumn{1}{c}{1} & \multicolumn{1}{c}{1} & \multicolumn{1}{c}{1} & \multicolumn{1}{c}{1} & \multicolumn{1}{c}{1} & \multicolumn{1}{c}{1} & \multicolumn{1}{c}{0} \\
			& \multicolumn{1}{l}{Wandering\_Line} & \multicolumn{1}{c}{1} & \multicolumn{1}{c}{0.933} & \multicolumn{1}{c}{0.875} & \multicolumn{1}{c}{0.933} & \multicolumn{1}{c}{0.923} & \multicolumn{1}{c}{0.933} & \multicolumn{1}{c}{0.045} \\
			\hline
			\multirow{3}{*}{3000 (Without Pre-training)} & Air\_Compressor&0.941&0.941&0.941&0.941&1&0.953&0.026\\
			& \multicolumn{1}{l}{Paired\_Doves}   & \multicolumn{1}{c}{1} & \multicolumn{1}{c}{1} & \multicolumn{1}{c}{1} & \multicolumn{1}{c}{1} & \multicolumn{1}{c}{1} & \multicolumn{1}{c}{1} & \multicolumn{1}{c}{0} \\
			& \multicolumn{1}{l}{Wandering\_Line} & \multicolumn{1}{c}{1} & \multicolumn{1}{c}{1} & \multicolumn{1}{c}{0.857} & \multicolumn{1}{c}{0.857} & \multicolumn{1}{c}{1} & \multicolumn{1}{c}{0.943} & \multicolumn{1}{c}{0.078} \\
			\hline
			\multirow{3}{*}{5000 (Without Pre-training)} & Air\_Compressor&1&0.941&0.941&0.941&0.941&0.953&0.026\\
			& \multicolumn{1}{l}{Paired\_Doves}   & \multicolumn{1}{c}{1} & \multicolumn{1}{c}{1} & \multicolumn{1}{c}{1} & \multicolumn{1}{c}{1} & \multicolumn{1}{c}{1} & \multicolumn{1}{c}{1} & \multicolumn{1}{c}{0} \\
			& \multicolumn{1}{l}{Wandering\_Line} & \multicolumn{1}{c}{1} & \multicolumn{1}{c}{1} & \multicolumn{1}{c}{1} & \multicolumn{1}{c}{1} &  \multicolumn{1}{c}{0.933} & \multicolumn{1}{c}{0.987} & \multicolumn{1}{c}{0.030} \\
			\hline
			\hline
			\multicolumn{9}{c}{With both data augmentation and pre-training}\\
			\hline
			\hline
			\multirow{3}{*}{500} & Air\_Compressor&0.941&0.941&0.941&0.941&1&0.953&0.026\\
			& \multicolumn{1}{c}{Paired\_Doves}   & \multicolumn{1}{c}{1} & \multicolumn{1}{c}{1} & \multicolumn{1}{c}{1} & \multicolumn{1}{c}{0.889} & \multicolumn{1}{c}{0.857} &
			\multicolumn{1}{c}{0.949} &\multicolumn{1}{c}{0.070}   \\
			& \multicolumn{1}{l}{Wandering\_Line} & \multicolumn{1}{c}{1} & \multicolumn{1}{c}{1} & \multicolumn{1}{c}{0.923} & \multicolumn{1}{c}{1} &  \multicolumn{1}{c}{1} & \multicolumn{1}{c}{0.985} & \multicolumn{1}{c}{0.034} \\
			\hline
			\multirow{3}{*}{2000} & Air\_Compressor&0.941&0.941&1&0.941&0.941&0.953&0.026\\
			& \multicolumn{1}{l}{Paired\_Doves}   & \multicolumn{1}{c}{1} & \multicolumn{1}{c}{1} & \multicolumn{1}{c}{1} & \multicolumn{1}{c}{1} &  \multicolumn{1}{c}{1} & \multicolumn{1}{c}{1} & \multicolumn{1}{c}{0}  \\
			& \multicolumn{1}{l}{Wandering\_Line} & \multicolumn{1}{c}{1} & \multicolumn{1}{c}{1} & \multicolumn{1}{c}{1} & \multicolumn{1}{c}{1} & \multicolumn{1}{c}{1} & \multicolumn{1}{c}{1} & \multicolumn{1}{c}{0}   \\
			\hline
			\multirow{3}{*}{3000} & Air\_Compressor&0.941&0.941&0.941&0.941&0.941&0.941&0\\
			& \multicolumn{1}{l}{Paired\_Doves}   & \multicolumn{1}{c}{1} & \multicolumn{1}{c}{1} & \multicolumn{1}{c}{1} & \multicolumn{1}{c}{1} & \multicolumn{1}{c}{1} & \multicolumn{1}{c}{1} & \multicolumn{1}{c}{0} \\
			& \multicolumn{1}{l}{Wandering\_Line} & \multicolumn{1}{c}{1} & \multicolumn{1}{c}{0.933} & \multicolumn{1}{c}{0.933} & \multicolumn{1}{c}{0.824} & \multicolumn{1}{c}{0.857} & \multicolumn{1}{c}{0.909} & \multicolumn{1}{c}{0.070}   \\
			\hline
			\multirow{3}{*}{5000} & Air\_Compressor&0.941&0.941&0.941&0.941&0.941&0.941&0\\
			& \multicolumn{1}{l}{Paired\_Doves}   & \multicolumn{1}{c}{1} & \multicolumn{1}{c}{1} & \multicolumn{1}{c}{1} & \multicolumn{1}{c}{1} & \multicolumn{1}{c}{1} & \multicolumn{1}{c}{1} & \multicolumn{1}{c}{0} \\
			& \multicolumn{1}{l}{Wandering\_Line} & \multicolumn{1}{l}{0.923} & \multicolumn{1}{c}{1} & \multicolumn{1}{c}{1} &  \multicolumn{1}{c}{0.923} &
			\multicolumn{1}{c}{1} & \multicolumn{1}{c}{0.969} &
			\multicolumn{1}{c}{0.042} \\
			\hline
			\hline
		\end{tabular}
		\label{table.5.3}%
	\end{table*}
	
	\begin{figure}
		\centering
		\includegraphics[scale=0.29]{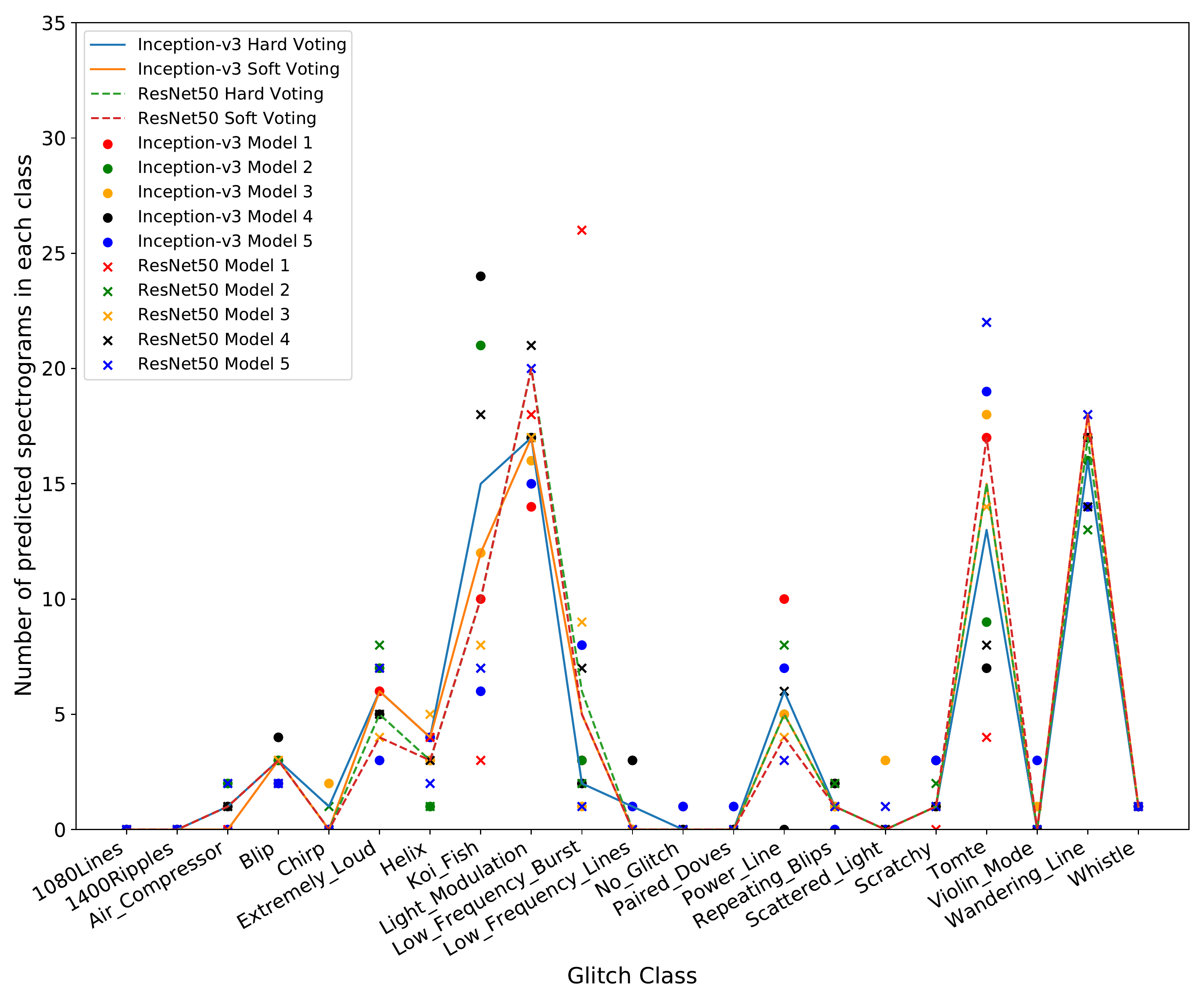}
		\caption{Predicted classes on the {\tt None\_of\_the\_Above} spectrograms with two different network architectures. For each network design, results from five individual model predictions, soft-voting and hard-voting are shown.}
		\label{fig.4.11}
	\end{figure}

We have also shown the confusion matrices for the Inception-V3 network architecture which attain the best performances among all the experiments without pre-training (Figure~\ref{fig.5.3}) and with pre-training (Figure~\ref{fig.5.4}). 
For the model that gives the optimal performance without pre-training and with $N_{total}=5000$ in data augmentation, perfect precision and recall have been achieved in ten classes ({\tt 1080lines}, {\tt 1400lines}, {\tt Chirp}, {\tt Extremely\_loud}, {\tt Helix}, {\tt Koi\_fish}, {\tt Paired\_doves}, {\tt Scattered\_Light},{\tt Scratchy} and {\tt Wandering\_line}). 
On the other hand, for the case with both pre-training and data augmentation ($N_{total}=2000$) incorporated, perfect precision and recall have been achieved in seven classes: ({\tt 1080lines}, {\tt Chirp}, {\tt Extremely\_loud}, {\tt Helix}, {\tt Koi\_fish}, {\tt Paired\_doves}, and {\tt Wandering\_line}).

	\begin{figure*}
		\centering
		\includegraphics[scale=0.85]{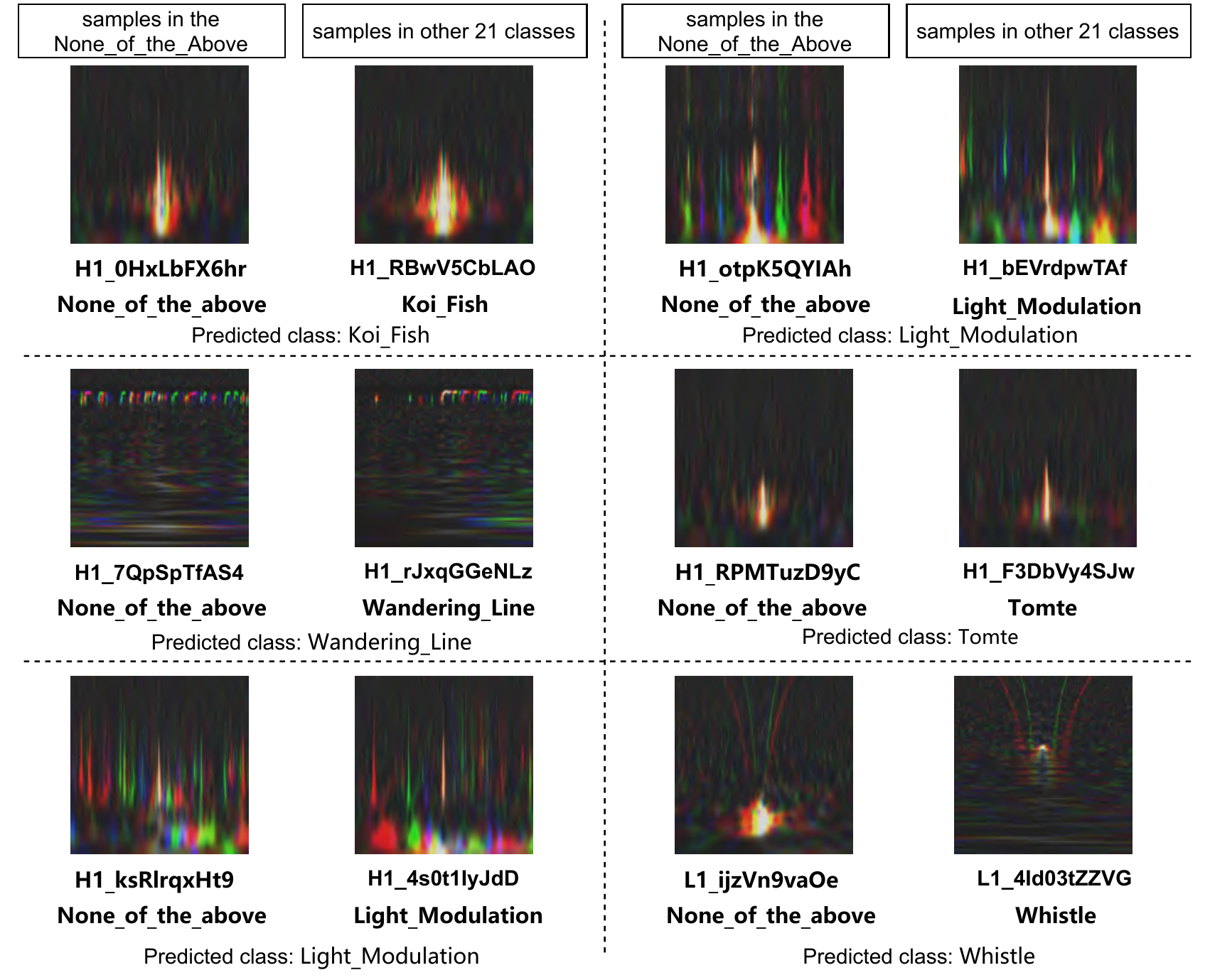}
		\caption{Comparison of some spectrograms in the {\tt None\_of\_the\_Above} class and some well-defined examples in the corresponding classes. The predictions of those spectrograms are assigned by our model. Designations of the selected data in {\it Gravity Spy} are given below each spectrogram.}
		\label{fig.x}
	\end{figure*}
	
	\subsection{Classifying spectrograms from {\tt None\_of\_the\_Above}}
The class {\tt None\_of\_the\_above} collects all the glitches that cannot be put into the other 21 classes without ambiguity. 
Since the spectrograms in this class are unlikely having a common origin, all the data in this class were ignored in all the aforementioned experiments. 
With the improved performance by using GAN-based data augmentation, we now attempt to classify the {\tt None\_of\_the\_Above} spectrograms into the other 21 well-defined classes. 

For predicting the classes of these unidentified spectrograms, we use the optimal models from two different network architectures in our experiment (i.e. Inception-V3 trained with pre-training and adopted $N_{total}=2000$, and ResNet50 trained with pre-training and adopted $N_{total}=500$). 
The reason of considering two network architectures is for cross-checking whether consistent predictions can be resulted from different network design.

Many methods are used for quantifying uncertainty in deep learning, such as Dropout \citep{srivastava2014dropout, gal2016dropout} and ensemble averaging methods \citep{lakshminarayanan2017simple, mani2017ensemble}. 
Since there is no Dropout layer in the ResNet50 network architecture \citep{he2016deep},we employed ensemble averaging method to quantify the uncertainty of the predictions in this experiment.
Similar to the other experiments in this work (see. Sec. 5.2), we repeat the prediction five times from each model and used both soft voting and hard voting for obtaining the final results from each network architecture. 
The distributions of the predicted classes from both networks are summarized in Figure~\ref{fig.4.11}. 
We show that different architecture designs and voting schemes lead to consistent results in this experiment.

To gain further confidence on the predictions, we have also visually compared the spectrograms from {\tt None\_of\_the\_Above} with those in the corresponding classes that our models have predicted. 
In Figure~\ref{fig.x}, we compare a few {\tt None\_of\_the\_Above} spectrograms with those from the classes which they are assigned by our model.
One should note the similarity between these unidentified spectrograms with the confirmed samples in their corresponding predicted classes.

\section{Summary \& Discussion}

In this work, we have demonstrated that the classification of spectrograms can be improved by incorporating GANs.  
A variant of GANs, ProGAN, is adopted in this experiment for generating high-resolution spectrograms. 
Here we summarized our major results:

\begin{enumerate}
\item
While conventional methods of data augmentation is not applicable in spectrogram classification, we have demonstrated that ProGAN not only provides a framework to generate spectrograms that are consistent with the real data (see Figure~\ref{fig.4.3}), but also increases the diversity which can result in a better generalization (Table~\ref{table.5.1}).
Using the high-resolution spectrograms generated by ProGAN, we enlarged the training data so as to uplifting both problems of small sample and imbalanced classes in the original data (cf. Algorithm~\ref{algorithm1}).
\item
We have shown that with the aid of GAN-base data augmentation, we can attain a performance of deep learning classification comparable with that obtained by the pre-trained model (Table~\ref{table.5.2} and Figure~\ref{fig.4.10}). 
This suggests that our framework can provide an alternative method of employing deep learning to classify gravitational wave spectrograms without using transfer-learning. 
\item
We found that the impact of our framework is particularly significant on the classes with small sample size. By including the GAN-generated spectrograms for training, the fluctuations in classifying the small classes can be greatly reduced (Table~\ref{table.5.3}). 
\end{enumerate}

Besides the glitch classification considered in this work, our method can be applied to a wider range of spectrograms obtained from the gravitational wave detectors. 
Different origins of gravitational wave signals show different patterns in the spectrograms. 
For example, a core collapsed supernova (CCSN) can trigger a number of possible features in the associated gravitational wave signals, e.g. a low-order g-mode from the proto-neutron star surface oscillation has its frequency grows in a timescale $<1$ s as it contracts. 
Such feature can be visualized as a defining pattern in the spectrograms. 
There is a general consensus that the outward-going shock will be stalled shortly after the rebounce. 
And it requires the neutrino heating from the proto-neutron star to re-accelerate the shock and drive the prompt explosion. 
This can trigger the standing accretion shock instability (SASI) which can generate the gravitational wave signal as a distinguishable feature in the spectrograms \citep[e.g see Fig. 9 \& 10 in ][]{Pan_2021}.

While the technique of matched filtering is found to be successful in uncovering the gravitational wave bursts from the compact binary coalescences which have well-defined waveforms as determined by numerical relativity, such method may have difficulties in searching the signals from CCSNe. 
Since the gravitational wave produced by CCSNe are expected to be affected by the turbulence, it should be stochastic in nature and hence the waveform cannot be deterministically predicted \citep[][]{Kotake_2009,10.1093/mnras/stz293}. 

Recently, the feasibility of detecting CCSN gravitational wave signals by deep learning has been considered \citep[e.g.][]{PhysRevD.102.043022}. 
However, because of the 3-D simulations of CCSNe are computationally expensive, the simulated data for training are suffered from the same problems that we have considered in this work \citep[i.e small sample size and imbalanced classes, see Table II in][for examples]{PhysRevD.102.043022}. 
By incorporating GAN-based data augmentation suggested in our work, not only the aforementioned problems can be relieved, the diversity of the training data can also be increased. 
Owing to the stochastic nature of the signals, the actual waveform from the real event is likely to show features which are unexpected from the simulations \citep{PhysRevD.102.043022}. 
By increasing the diversity of the training data and hence enlarging the parameter space coverage, GANs can potentially enhance the detectability of such signals. 

\section*{Acknowledgements}
CYH is supported by the research fund of Chungnam National University. 
The work of JY is also supported by  the National Research Foundation of Korea through grants 2016R1A5A1013277 and 
2022R1F1A1073952. In addition,
the work of Jianqi Yan and Alex Po Leung is supported by the Science and Technology Development Fund, Macau SAR (No. 0079/2019/A2).
	
	%%%%%%%%%%%%%%%%%%%%%%%%%%%%%%%%%%%%%%%%%%%%%%%%%%
\section*{Data Availability}
The data underlying this article are available in Zenodo, at \url{https://zenodo.org/record/1476156#.YUSN6bgzaUn}. DOI: \url{https://dx.doi.org/10.5281/zenodo.1476156}.

%	The inclusion of a Data Availability Statement is a requirement for articles published in MNRAS. Data Availability Statements provide a standardised format for readers to understand the availability of data underlying the research results described in the article. The statement may refer to original data generated in the course of the study or to third-party data analysed in the article. The statement should describe and provide means of access, where possible, by linking to the data or providing the required accession numbers for the relevant databases or DOIs.

	
	
	\bibliographystyle{mnras}
	\bibliography{bibliography} % if your bibtex file is called example.bib

	%\begin{thebibliography}{99}
	%\end{thebibliography}
	
	%%%%%%%%%%%%%%%%%%%%%%%%%%%%%%%%%%%%%%%%%%%%%%%%%%
	
	%%%%%%%%%%%%%%%%% APPENDICES %%%%%%%%%%%%%%%%%%%%%
	
\appendix

\section{Robustness test of t-SNE experiments with different random seeds}
\begin{figure*}
	\centering
	\includegraphics[scale=0.45]{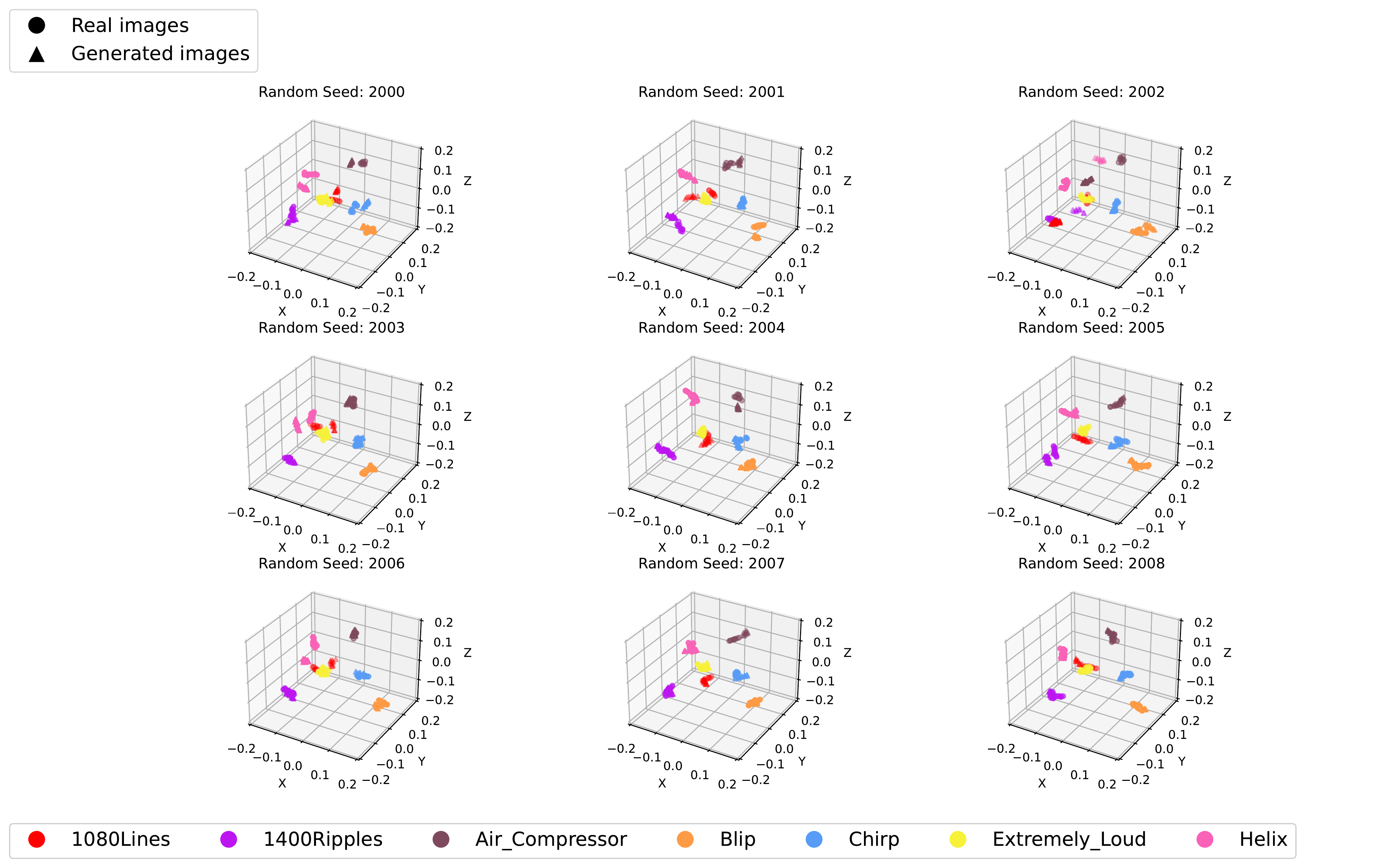}
	\caption{\textbf{t-SNE experimental results with different random seeds. Seven types of glitches in the experiment are selected.}}
	\label{appendix_A}
\end{figure*}%

	%%%%%%%%%%%%%%%%%%%%%%%%%%%%%%%%%%%%%%%%%%%%%%%%%%

	% Don't change these lines
	\bsp	% typesetting comment
	\label{lastpage}
\end{document}